\documentclass[12pt,aps,prd,preprint,tightenlines,superscriptaddress,
amsmath,amssymb,nofootinbib]{revtex4-1} 

\RequirePackage[colorlinks=true
,urlcolor=blue
,anchorcolor=blue
,citecolor=blue
,filecolor=blue
,linkcolor=blue
,menucolor=blue
,pagecolor=blue
,linktocpage=true
,pdfproducer=medialab
,pdfa=true
]{hyperref}

\allowdisplaybreaks

\usepackage{amsmath,amssymb,amsthm,amsfonts}
\usepackage{graphicx}
\usepackage{dcolumn}	
\usepackage{hyperref}      	
\usepackage{setspace}
\usepackage[usenames, dvipsnames]{color}
\usepackage{slashed}
\usepackage{enumitem}
\usepackage{siunitx}
\usepackage{multirow}

\newcommand{\PRE}[1]{{#1}} 

\newcommand{\be}{\begin{equation}\begin{aligned}}
\newcommand{\ee}{\end{aligned}\end{equation}}
\newcommand{\beq}{\begin{equation}}
\newcommand{\eeq}{\end{equation}}
\newcommand{\beqa}{\begin{eqnarray}}
\newcommand{\eeqa}{\end{eqnarray}}

\newcommand{\ifb}{\text{fb}^{-1}}
\newcommand{\iab}{\text{ab}^{-1}}

\newcommand{\mev}{\text{MeV}}
\newcommand{\gev}{\text{GeV}}
\newcommand{\tev}{\text{TeV}}
\newcommand{\fb}{\text{fb}}

\newcommand{\micm}{\mu\text{m}}
\newcommand{\mm}{\text{mm}}
\newcommand{\cm}{\text{cm}}
\newcommand{\m}{\text{m}}

\newcommand{\murad}{\mu\text{rad}}
\newcommand{\g}{\text{g}}
\newcommand{\kg}{\text{kg}}

\newcommand{\ns}{\text{ns}}

\newcommand{\s}{\text{s}}

\renewcommand{\eqref}[1]{Eq.~(\ref{#1})}

\newcommand{\secref}[1]{Sec.~\ref{sec:#1}}
\newcommand{\secsref}[2]{Secs.~\ref{sec:#1} and \ref{sec:#2}}

\newcommand{\figref}[1]{Fig.~\ref{fig:#1}}

\newcommand{\Figref}[1]{Figure~\ref{fig:#1}}
\newcommand{\tableref}[1]{Table~\ref{table:#1}}

\newcommand{\red}[1]{\textcolor{red}{#1}}

\RequirePackage[normalem]{ulem}

\begin{document}

\preprint{Submitted to the LHCC, 18 July 2018 \hspace*{1.4in}  CERN-LHCC-2018-030, LHCC-I-032}
\preprint{UCI-TR-2018-18, KYUSHU-RCAPP-2018-05}

\title{ \vspace*{.2in}
LETTER OF INTENT \\
\vspace*{.3in}
{\Large FASER} \\ \vspace*{.1in}
{\Large FORWARD SEARCH EXPERIMENT AT THE LHC}
\PRE{\vspace*{.3in}}}

\author{Akitaka Ariga}
\affiliation{Universit\"at Bern, Sidlerstrasse 5, CH-3012 Bern, Switzerland}

\author{Tomoko Ariga}
\affiliation{Universit\"at Bern, Sidlerstrasse 5, CH-3012 Bern, Switzerland}
\affiliation{Kyushu University, Nishi-ku, 819-0395 Fukuoka, Japan}

\author{Jamie Boyd}
\email[Contact email: ]{Jamie.Boyd@cern.ch}
\affiliation{CERN, CH-1211 Geneva 23, Switzerland}

\author{David~W.~Casper}
\affiliation{Department of Physics and Astronomy, 
University of California, Irvine, CA 92697-4575, USA}

\author{Jonathan~L.~Feng}
\email[Contact email: ]{jlf@uci.edu}
\affiliation{Department of Physics and Astronomy, 
University of California, Irvine, CA 92697-4575, USA}

\author{Iftah Galon}
\affiliation{New High Energy Theory Center, Rutgers, The State University of New Jersey, Piscataway, New Jersey 08854-8019, USA}

\author{Shih-Chieh Hsu}
\affiliation{University of Washington, PO Box 351560, Seattle, WA 98195-1560, USA}

\author{Felix Kling}
\affiliation{Department of Physics and Astronomy, 
University of California, Irvine, CA 92697-4575, USA}

\author{Hidetoshi Otono}
\affiliation{Kyushu University, Nishi-ku, 819-0395 Fukuoka, Japan}

\author{Brian Petersen}
\affiliation{CERN, CH-1211 Geneva 23, Switzerland}

\author{Osamu Sato}
\affiliation{Nagoya University, Furo-cho, Chikusa-ku, Nagoya-shi 464-8602, Japan}

\author{Aaron~M.~Soffa}
\affiliation{Department of Physics and Astronomy, 
University of California, Irvine, CA 92697-4575, USA}

\author{Jeffrey~R.~Swaney}
\affiliation{Department of Physics and Astronomy, 
University of California, Irvine, CA 92697-4575, USA}

\author{Sebastian Trojanowski\vspace*{.2in}}
\affiliation{\mbox{National Centre for Nuclear Research, Ho{\. z}a 69, 00-681 Warsaw, Poland}\vspace*{.4in}}


\begin{abstract}
\PRE{\vspace*{.2in}}
FASER is a proposed small and inexpensive experiment designed to search for light, weakly-interacting particles at the LHC.  Such particles are dominantly produced along the beam collision axis and may be long-lived, traveling hundreds of meters before decaying.  To exploit both of these properties, FASER is to be located along the beam collision axis, 480 m downstream from the ATLAS interaction point, in the unused service tunnel TI18.  We propose that FASER be installed in TI18 in Long Shutdown 2 in time to collect data from 2021-23 during Run 3 of the 14 TeV LHC.  FASER will detect new particles that decay within a cylindrical volume with radius $R= 10~\cm$ and length $L = 1.5~\m$.  With these small dimensions, FASER will complement the LHC's existing physics program, extending its discovery potential to a host of new particles, including dark photons, axion-like particles, and other CP-odd scalars.  A FLUKA simulation and analytical estimates have confirmed that numerous potential backgrounds are highly suppressed at the FASER location, and the first {\em in situ} measurements are currently underway.  We describe FASER's location and discovery potential, its target signals and backgrounds, the detector's layout and components, and the experiment's preliminary cost estimate, funding, and timeline. 
\end{abstract}

\pagenumbering{roman}
\maketitle
\tableofcontents
\clearpage

\pagenumbering{arabic}

\section{Introduction}
\label{sec:introduction}

For decades, the leading examples of new physics targets at particle colliders were particles with TeV-scale masses and ${\cal O}(1)$ couplings to the standard model (SM).  More recently, however, there is a growing and complementary interest in new particles that are much lighter and more weakly coupled~\cite{Battaglieri:2017aum}.  Among their many motivations, such particles may yield dark matter with the correct thermal relic density and resolve outstanding discrepancies between theory and low-energy experiments~\cite{Bennett:2006fi, Pohl:2010zza, Krasznahorkay:2015iga}.  Perhaps most importantly, new particles that are light and weakly coupled can be discovered by relatively inexpensive, small, and fast experiments with potentially revolutionary implications for particle physics and cosmology.

If new particles are light and very weakly coupled, the focus at the LHC on particle searches at high transverse momentum ($p_T$) may be completely misguided.  In contrast to TeV-scale particles, which are produced more or less isotropically, light particles with masses in the MeV to GeV range are dominantly produced at low $p_T \sim 100~\mev - \gev$.  In addition, because the new particles are extremely weakly coupled, very large SM event rates are required to discover the rare new physics events.  These rates are available, not at high $p_T$, but at low $p_T$: at the 13 TeV LHC, the total inelastic $pp$ scattering cross section is $\sigma_{\text{inel}}(13~\text{TeV}) \approx 75~\text{mb}$~\cite{Aaboud:2016mmw, VanHaevermaet:2016gnh}, with most of it in the very forward direction. In upcoming runs at 14 TeV, where the inelastic cross section is very similar, we expect
\be
N_{\text{inel}} \approx 1.1 \times 10^{16} \ (2.2 \times 10^{17})
\label{eq:ppcollisions}
\ee
inelastic $pp$ scattering events for an integrated luminosity of $150~\text{fb}^{-1}$ at LHC Run 3 ($3~\text{ab}^{-1}$ at the HL-LHC). Even extremely weakly-coupled new particles may therefore be produced in sufficient numbers in the very forward region.  Given their weak coupling to the SM, such particles are typically long-lived and travel a macroscopic distance before decaying back into SM particles.  Moreover, such particles may be highly collimated.  For example, new particles that are produced in pion or $B$ meson decays are typically produced within angles of $\theta \sim \Lambda_{\text{QCD}} / E$ or $m_B / E$ of the beam collision axis, where $E$ is the energy of the particle.  For $E \sim \text{TeV}$, this implies that even $\sim 500~\m$ downstream, such particles have only spread out $\sim 10~\cm - 1~\m$ in the transverse plane.  A small and inexpensive detector placed in the very forward region may therefore be capable of extremely sensitive searches. 

FASER~\cite{Feng:2017uoz}, the ForwArd Search ExpeRiment, is specifically designed to take advantage of this opportunity.  An ideal location exists in TI18\footnote{Note added: Since this document was prepared, additional measurements made by the CERN survey team during LHC Technical Stop 2 in September 2018 have shown that tunnel TI12 on the other side of LHC interaction point IP1 will more easily accommodate the detector with the dimensions presented here.  TI12 is the same distance from IP1 as TI18 and the expected signal and backgrounds are very similar.  FASER is now expected to be located in TI12.}, an existing and unused side tunnel that is 480 m downstream from the ATLAS interaction point (IP). We propose that FASER be installed along the beam collision axis in TI18 in Long Shutdown 2 (LS2) from 2019-20 in time to collect data in Run 3 from 2021-23. With an active volume of only $0.16~\m^3$, FASER will complement the LHC's existing physics program, with remarkable sensitivity to dark photons, axion-like particles, and other proposed particles. In the following sections, we discuss FASER's location and discovery potential, the detector's layout and components, backgrounds, and the experiment's preliminary cost estimate, funding, and timeline.

\section{Detector Location}
\label{sec:location}

\begin{figure}[t]
\centering
\includegraphics[width=0.99\textwidth]{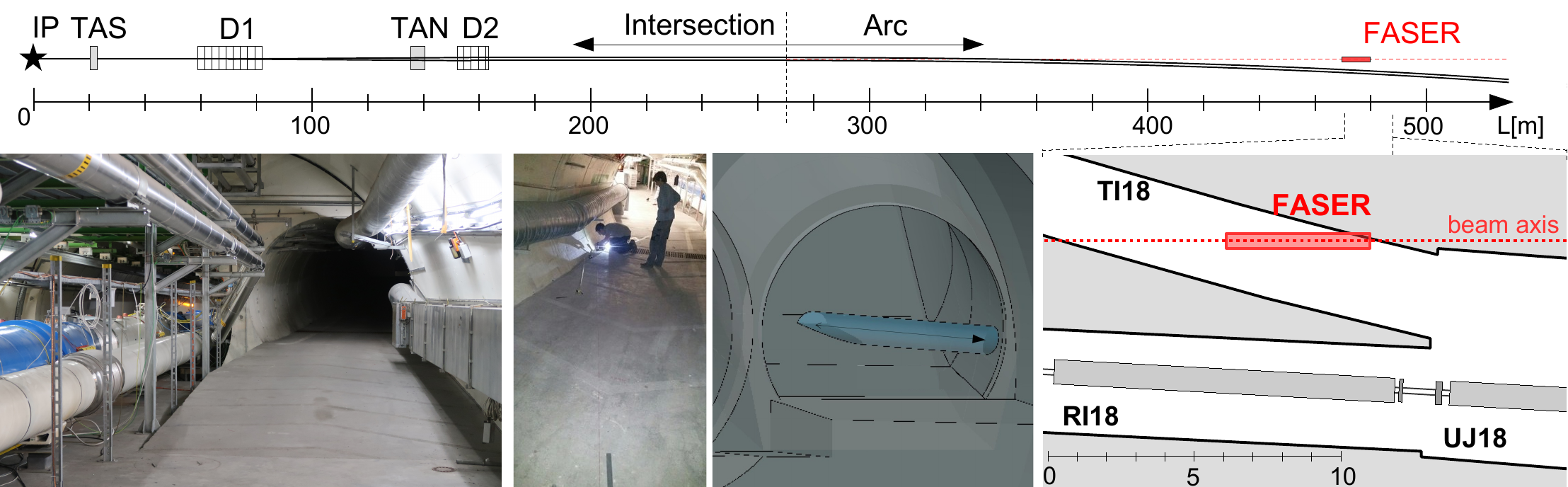} 
\caption{
Location of FASER. Top panel: A schematic drawing of the LHC and the very forward infrastructure downstream from the ATLAS IP, with FASER located 480 m from the IP, after the LHC tunnel starts to curve. Bottom panels, left to right: the view toward the ATLAS IP, with the main LHC tunnel on the left and tunnel TI18 on the right; the beam collision axis marked on the floor of TI18 by the CERN survey team; the location of the beam collision axis in TI18; and a map showing the intersection of the beam collision axis and TI18 from above.
} 
\label{fig:Infrastructure}
\end{figure}

As shown in Fig.~\ref{fig:Infrastructure}, an ideal location for FASER is along the beam collision axis, 480 meters downstream from the ATLAS IP in service tunnel TI18. This tunnel was formerly used to connect the SPS to the LEP tunnel, but it is currently empty and unused. Light, weakly-interacting particles produced at the ATLAS IP will travel along the beam collision axis through matter without interacting, and then can decay in TI18.  

In this location, then, FASER harnesses the enormous, previously ``wasted,'' cross section for very forward physics to search for light, weakly-coupled new particles.  This location also benefits from the fact that, when long-lived particles (LLPs) are produced at the unprecedented center-of-mass energies of the LHC, their large boosts result in decays that are far beyond the main LHC infrastructure in regions where backgrounds are highly suppressed. 

In more detail, as shown in the bottom panels of \figref{Infrastructure}, the beam collision axis emerges from the floor of TI18 for a distance of roughly 3.8 m before intersecting the side wall.  The beam collision axis has been located to within a mm by the CERN survey team, but at any given time, its precise location depends on the beam crossing angle at ATLAS.  For example, the current beam crossing half angle of $150~\mu\text{rad}$ in the upward direction raises the beam collision axis by 7 cm in TI18 relative to what is shown in \figref{Infrastructure}. Prior to installing FASER, we propose that the floor in TI18 be lowered by 50 cm to allow a longer detector to be placed along the beam collision axis.  This will not disrupt essential services, and no other excavation is required. Detailed studies are ongoing to assess exactly how long the detector can be for the different beam crossing angles currently envisioned for Run 3, but preliminary estimates suggest a length of 5 m will be possible.

\section{New Physics Discovery Potential}
\label{sec:DiscoveryPotential}

The potential for discovering new LLPs has been studied for a plethora of new physics models~\cite{Feng:2017uoz, Feng:2017vli, Batell:2017kty, Kling:2018wct, Helo:2018qej, Bauer:2018onh, Feng:2018noy, Berlin:2018jbm}, both for FASER and for a possible larger follow-up experiment, FASER 2, which would run in the HL-LHC era. To illustrate this discovery potential, here we present results for three examples of light, weakly-interacting new particles: dark photons~\cite{Feng:2017uoz}, axion-like particles (ALPs)~\cite{Feng:2018noy}, and CP-odd scalars that couple dominantly to SM fermions. 

\textbf{Dark photons:} A massive dark photon arises when a hidden sector contains a broken $U(1)$ gauge symmetry. The hidden sector's gauge boson can then mix with the SM photon via kinetic mixing and obtain a small coupling to the SM electromagnetic current proportional to the kinetic mixing parameter $\epsilon$, leading to the Lagrangian terms
\be 
\mathcal{L} \supset \frac{1}{2} \red{m}_{\red{A'}}^{2} A'^2 -  \red{\epsilon} \, e j_{EM}^\mu A_{\mu}' \ ,
\ee
where $\epsilon$ is naturally small if the mixing is loop-induced.  Dark photons are primarily produced in the decay of light mesons or via dark bremsstrahlung and are therefore very collimated around the beam collision axis. They can decay into all kinematically-allowed charged particles. In the parameter space probed at FASER, they decay via $A' \to e^+ e^-$, and most of the signal is confined to within 10 cm of the beam collision axis~\cite{Feng:2017uoz}.

\textbf{Axion-like particles (ALPs):} ALPs are pseudoscalar SM-singlets that appear as pseudo-Nambu-Goldstone bosons in theories with broken global symmetries. We consider a low-energy effective theory in which an ALP couples only to photons through the dimension-5 interaction, leading to the Lagrangian terms 
\be 
\mathcal{L} \supset - \frac{1}{2} \red{m}_{\red{a}}^2 a^2 - \frac{1}{4} \red{g_{a\gamma\gamma}} \, a F^{\mu\nu} \widetilde{F}_{\mu\nu} \ .
\ee
ALPs reaching FASER are predominantly produced through the Primakoff process by high energy photons colliding with the TA(X)N~\cite{Feng:2018noy}. The initial photons are highly collimated along the beam axis, leading to similarly collimated signal when ALPs decay via $a \to \gamma \gamma$.

\textbf{CP-odd scalars:} For our last example, we consider light pseudoscalars that couple dominantly to SM fermions. Such particles could be ALPs or part of an extended Higgs sector. Following the model presented in Ref.~\cite{Dolan:2014ska}, we require the fermion couplings to be proportional to Yukawa couplings, leading to the Lagrangian terms 
\be 
\mathcal{L} \supset - \frac{1}{2} \red{m}_{\red{A}}^2 A^2 - i \sum_f \red{g_{Aff}} \, y_f  \, A \bar{f} \gamma^5 f \ .
\ee
These LLPs are mainly produced in the heavy quark decay $b \to s A$, leading to a larger spread around the beam collision axis. At FASER, the leading signal is from $A \to \mu^+ \mu^-$.

\begin{figure}[tbp]
\centering
\includegraphics[width=0.32\textwidth]{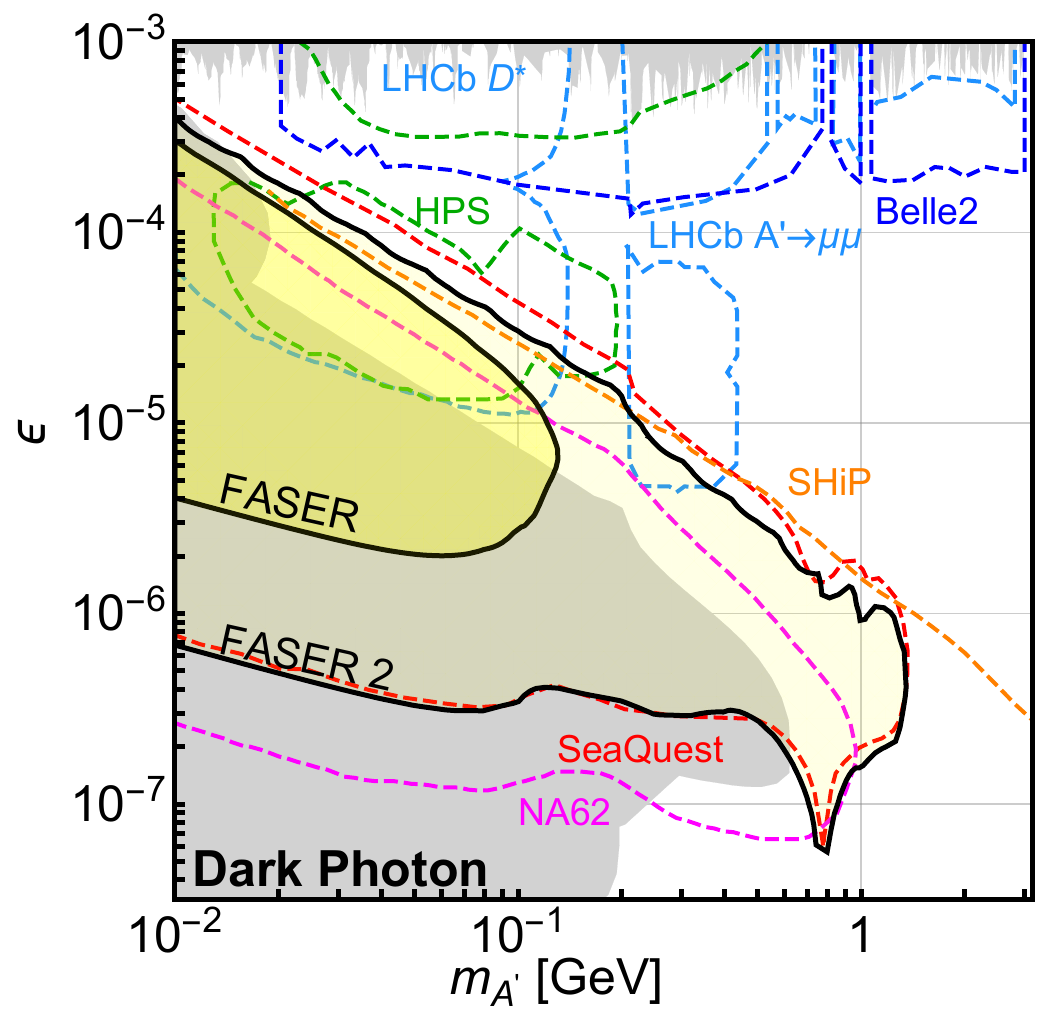} 
\includegraphics[width=0.32\textwidth]{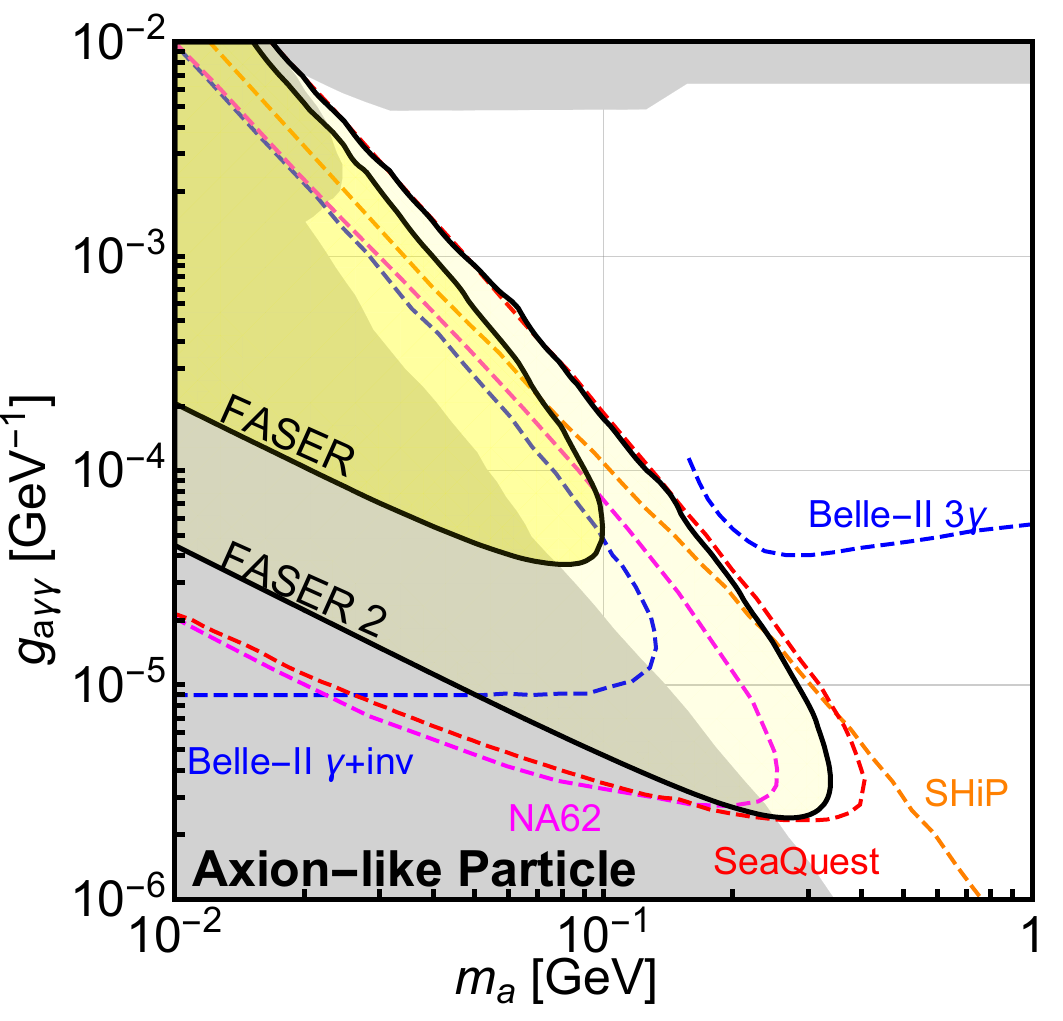}
\includegraphics[width=0.32\textwidth]{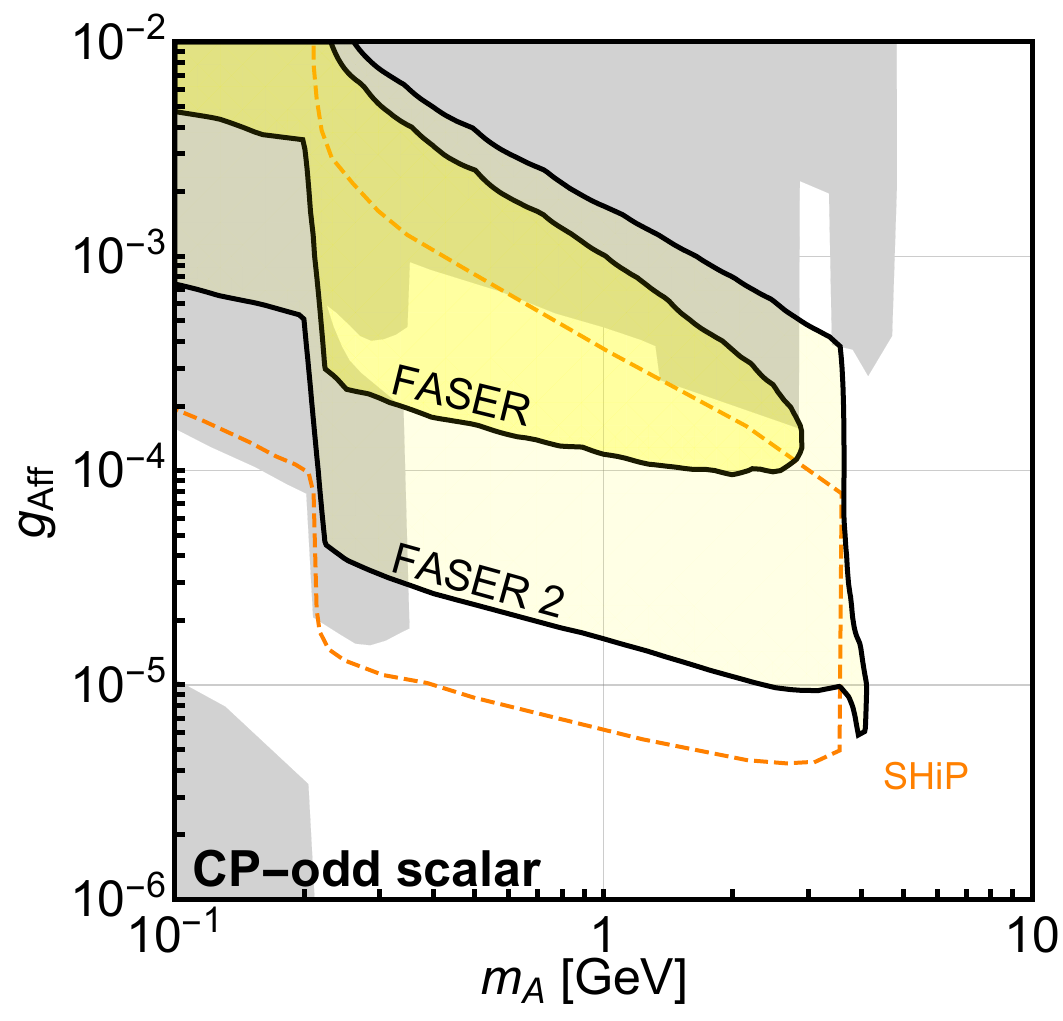} 
\caption{
Projected sensitivity reaches for FASER at the 14 TeV LHC Run 3 with $150~\ifb$ for dark photons (left), axion-like particles (center), and CP-odd scalars (right). The gray-shaded regions are excluded by current bounds. For comparison we also show the sensitivities of FASER 2, a possible upgraded detector running in the HL-LHC era (see text), and other current and proposed experiments: NA62 assumes $3.9 \times 10^{17}$ protons on target (POT) while running in a beam dump mode that is being considered for LHC Run 3~\cite{Dobrich:2015jyk}; SeaQuest assumes $1.44 \times 10^{18}$ POT, which could be obtained in two years of parasitic data taking and requires additionally the installation of a calorimeter~\cite{Berlin:2018pwi, Bauer:2018onh}; the proposed beam dump experiment SHiP assumes $\sim 2 \times 10^{20}$ POT collected in 5 years of operation~\cite{Alekhin:2015byh, Dobrich:2015jyk}; Belle-II and LHCb assume the full expected integrated luminosity of $50~\iab$~\cite{Dolan:2017osp} and $300~\ifb$~\cite{Ilten:2015hya, Ilten:2016tkc}, respectively; and HPS assumes 4 weeks of data at JLab at each of several different beam energies~\cite{Moreno:2013mja,Battaglieri:2017aum}. 
}
\label{fig:Reach}
\end{figure}

\Figref{Reach} shows FASER's sensitivity reach for each of these three models.  These results are for a cylindrical decay volume of radius $R=10~\cm$ and length $L=1.5~\m$ at the 14 TeV LHC with $150~\ifb$.  They are $N = 3$ signal event contours and so assume 100\% signal efficiency and negligible background (see \secsref{requirements}{bkg}).  We see that even with such a small active decay volume, FASER can probe significant new regions of parameter space in a variety of models. These results use the EPOS-LHC~\cite{Pierog:2013ria} Monte Carlo generator, which is tuned to the recently available forward scattering data from the LHC~\cite{N.Cartiglia:2015gve}, to simulate forward light meson production and FONLL~\cite{Cacciari:2012ny} with CTEQ~6.6 to simulate heavy meson production. For comparison, we also show the projected reach of other proposed experiments.  For the CP-odd scalar model, results for the proposed LHC experiments MATHUSLA~\cite{Curtin:2018mvb} and CODEX-b~\cite{Gligorov:2017nwh} can be expected to be complementary and probe lower couplings than FASER. We also show the reach of FASER 2, a possible larger detector with a decay volume of radius $R=1~\m$ and length $L=5~\m$ collecting data at the 14 TeV HL-LHC with $3~\iab$. Such an upgrade would extend FASER's reach significantly, particularly towards larger masses. 

The general features of the sensitivity curves can be understood as follows: for relatively large $\epsilon$, the sensitivity is reduced because the LLPs tend to decay before they reach FASER.  The reach is extremely sensitive to $\epsilon$, and changing other parameters, for example, requiring 10 signal events instead of 3, or including a 50\% signal efficiency factor, leads to almost imperceptible changes in the sensitivity reach contours.  In contrast, for relatively small $\epsilon$ and large LLP masses, the reach is limited by the LLP production cross section, and larger datasets can extend the reach in parameter space significantly.  

The regions of parameter space probed by FASER are of interest for both particle physics and cosmology. For example, if a dark photon couples to a dark matter particle with mass $\sim m_{A'}$, the dark matter can have the correct thermal relic density if $m_{A'}  \sim \epsilon m_{\text{weak}}$, where $m_{\text{weak}} \sim 1~\tev$. For $m_{A'} \sim 10-100~\mev$, one obtains $\epsilon \sim 10^{-5} - 10^{-4}$, which is a region of parameter space that will be probed by FASER.

Finally, we note that FASER's physics potential is not restricted to the models mentioned above. Other particles probed by FASER and FASER 2 include dark Higgs bosons~\cite{Feng:2017vli}, flavor-specific scalar mediators~\cite{Batell:2017kty}, heavy neutral leptons~\cite{Kling:2018wct,Helo:2018qej}, $R$-parity violating neutralinos~\cite{Helo:2018qej}, $U(1)_{B-L}$ gauge bosons~\cite{Bauer:2018onh}, and inelastic dark matter~\cite{Berlin:2018jbm}.

\section{Detector Overview}
\label{sec:signal}

\subsection{Signal and Background: General Characteristics}
\label{sec:general}

FASER will search for LLPs that are produced at or close to the IP, move along the beam collision axis, and decay visibly within FASER. The characteristic event is
\be
  p p  \to \text{LLP} +X, \quad  \text{LLP travels} \ \sim 480~\text{m}, \quad \text{LLP} \to e^+ e^- , \mu^+ \mu^- , \pi^+ \pi^-, \gamma \gamma, \ldots
\ee
LLPs that travel in the very forward direction and decay in FASER typically have very high energies $\sim~\text{TeV}$.  The target signal at FASER is therefore striking: two oppositely charged tracks or two photons with $\sim \tev$ energies that emanate from a common vertex inside the detector and have a combined momentum that points back through 90 m of rock to the IP. 

When the LLPs decay, because they are light and highly boosted, their decay products are very collimated.  For example, for an LLP with mass $m=100~\mev$ and energy $E=1~\tev$, the typical opening angle of the decay products is $\theta \sim m/E \sim 100~\murad$ implying a separation of only $\sim 100~\micm$ after traveling $1~\m$. To use these striking kinematic features to distinguish signal from background, a measurement of the two individual decay products is highly desirable. 

For charged tracks, a magnetic field of $0.5~\text{T}$, achievable with permanent magnets, is able to both split the tracks sufficiently and allow for a track momentum measurement, as discussed in \secref{perfTracker}. Tracking layers, surrounded by a magnet to separate highly collimated tracks, and supplemented by a calorimeter to distinguish electrons from muons and provide additional energy measurements, will be the key components of FASER. 

For the di-photon signal, distinguishing the two photons requires a calorimeter with exquisite spatial resolution.  For this purpose, a pre-shower detector to convert and spatially resolve the photons is under consideration.  As discussed in \secref{bkg}, however, the expected background of single, $\sim \tev$ photons is very low, and so even if di-photon events are mis-reconstructed as single showers, they may be indicative of new physics.

The natural (rock) and LHC infrastructure (magnets and absorbers) shielding eliminates most potential backgrounds. Muons and neutrinos are the only known particles that can transport TeV energies through 90 m of rock between the IP and FASER. The dominant source of background is radiative processes associated with muons from the IP, which is identified by the presence of a high-energy muon traversing the full detector. This is suppressed by using a charged particle veto layer at the front of the detector. Additional backgrounds from neutrino interactions within the detector are small and generally have different kinematics. 

\subsection{Detector Layout}
\label{sec:layout}

The detector design is driven by the following considerations:
\begin{itemize}[leftmargin=0.16in]
\setlength{\itemsep}{-0.03in}
\item the detector should be highly sensitive to the signals discussed above and allow multiple background estimates from the data; 
\item the active area of the detector should lie close to the floor so that it can  intersect the beam collision axis for all possible beam crossing angles; 
\item the detector should be inexpensive and robust, using well-established technologies and, where possible, existing detector components;
\item the required services (power, cooling, gas, etc.) should be minimized, since access to the detector will not be possible when the LHC is running; and
\item the detector components are limited by the need to transport them through the LHC tunnel and over the LHC dipoles at the entrance to TI18.
\end{itemize}

The layout of the proposed FASER detector is illustrated in \figref{DetectorLayout}. At the entrance to the detector, a double layer of scintillators is used to veto charged particles coming through the cavern wall from the IP, primarily high-energy muons. In between the layers is a 20-radiation-lengths-thick layer of lead for converting any photons produced in the wall into electromagnetic showers that can be efficiently vetoed by the scintillators.

\begin{figure}[t]
\centering
\includegraphics[width=0.87\textwidth]{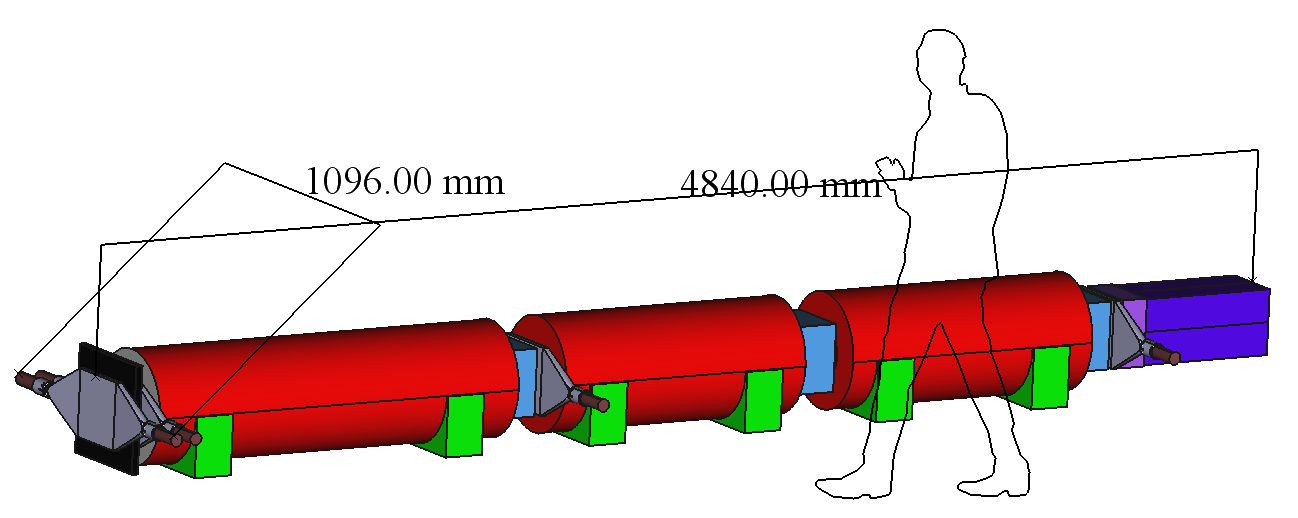} 
\caption{
Layout of the proposed FASER detector. LLPs enter from the left.  The detector components include scintillators (gray), dipole magnets (red), tracking stations (blue), a pre-shower detector (light purple), and a calorimeter (dark purple).
}
\label{fig:DetectorLayout}
\end{figure}

The veto layer is followed by a $1.5~\m$ long, 0.5 T permanent dipole magnet with a $10~\cm$ aperture radius. This is the decay volume for LLPs decaying into a pair of charged particles, with the magnet providing a horizontal kick to separate the decay products to a detectable distance. The decay volume is not foreseen to be under vacuum.

After the decay volume is a spectrometer consisting of two $1~\m$ long, 0.5 T dipole magnets with three tracking stations, each composed of layers of precision silicon strip detectors, located at either end and in between the magnets. The magnet covering the decay volume, and those in the spectrometer, will have their fields aligned to give the maximum separation for charged particles in the bending plane. Scintillator planes for triggering and precision time measurements are located at the entrance and exit of the spectrometer. The primary purpose of the spectrometer is to observe the characteristic signal of two oppositely charged particles pointing back towards the IP, measure their momenta, and sweep out low-momentum charged particles before they reach the final layer of the spectrometer. 

The final component is the electromagnetic calorimeter. This will identify high-energy electrons and photons and measure the total electromagnetic energy. As the primary signals are two close-by electrons or photons, these cannot be resolved by the main calorimeter. It is therefore under consideration to place a high-granularity pre-shower detector between the last tracking layer and the main calorimeter. This could be constructed from a layer of tungsten and one or two layers of silicon strip detectors.

\section{Detector Requirements, Optimization and Performance}
\label{sec:requirements}

The detector is designed to identify the two high-momentum, oppositely-charged particles from LLP decay and reject backgrounds that are topologically or kinematically inconsistent with the expected signal.

FASER's very high energy threshold for analysis is a powerful background rejection tool. Conservatively requiring $A'$ decay products above $100~\gev$ introduces negligible loss of physics sensitivity, and virtually eliminates non-instrumental SM backgrounds. To identify the signal with single-event sensitivity, the detector must be able to:
\begin{itemize}[leftmargin=0.16in]
\setlength{\itemsep}{-0.03in}
\item efficiently tag charged particles entering from the IP direction with energies $> 100~\gev$,
\item locate and distinguish exactly two oppositely-charged, nearly collinear primary particles (both with energies above $100~\gev$) consistent with an origin inside the detector decay volume and the expected direction, and
\item confirm the expected high-energy, electromagnetic character of the signal by robust, independent means.
\end{itemize}

\subsection{Tracker}
\label{sec:perfTracker}

The tracker's performance is constrained by limited space, services, and budget, as well as the high energy of the signal. The minimum requirement is the ability to distinguish two tightly collimated high-momentum charged tracks. If additional primary tracks are present, the tracker functions as a topological veto.  A magnetic field is applied to the decay volume, even though no measurements are made there, to increase the spatial separation of tracks before they reach the first tracker plane.

\Figref{tracking} shows the expected track separation in the forward (planes 1 and 2) and central (planes 3, 4, 5, and 6) tracking stations for $m_{A'}=100~\mev$. There is no significant dependence on the dark photon mass. The dotted vertical line corresponds to an expected one/two-track separation threshold of $300~\micm$. For $E_{A'} > 2~\tev$, the oppositely-charged tracks are typically separated by less than this distance at the first tracking station, but the majority are sufficiently separated by the second for all energies. 

\begin{figure}[tbp]
\centering
\includegraphics[width=0.38\textwidth]{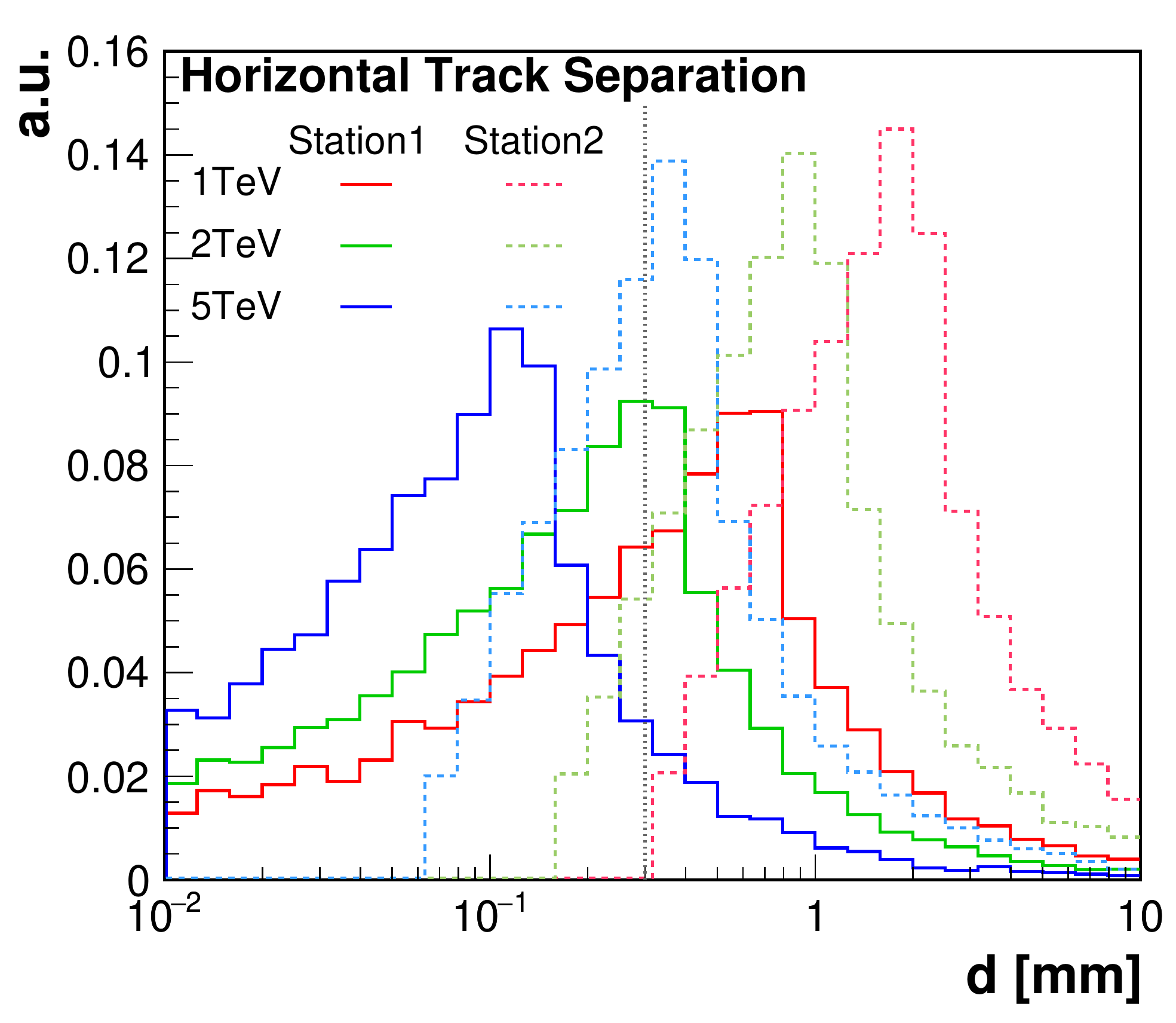} \qquad
\includegraphics[width=0.38\textwidth]{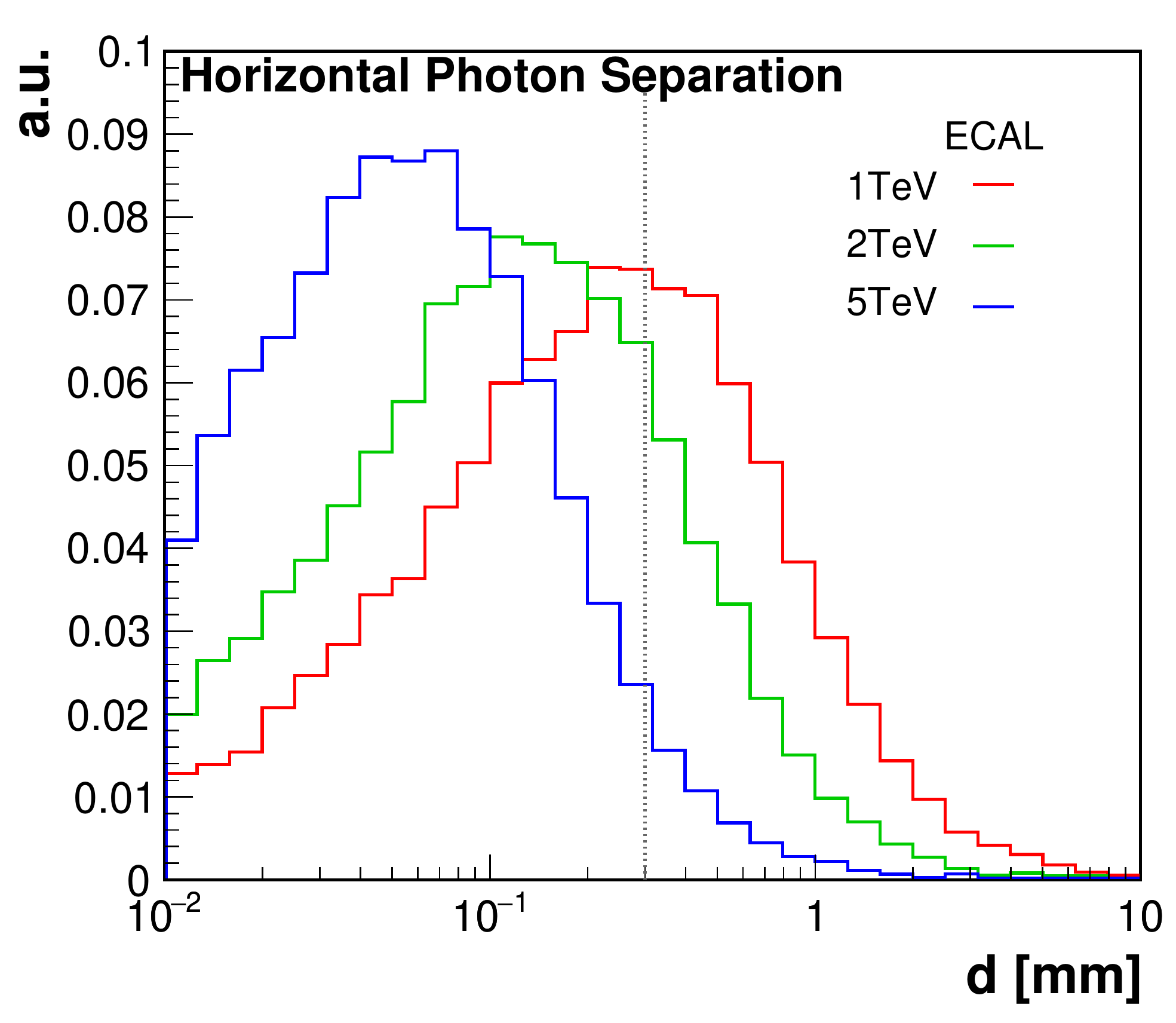} 
\caption{ 
Left: Separation of charged tracks from decay of a dark photon ($m_{A'}=100~\mev$) at three energies.  The solid (dashed) histograms are the expected separation in the forward (central) tracking stations, as defined in the text. Right: Separation of photons from the decay of an ALP ($m_{a}=100~\mev$) at the end of the tracking system. In both panels, the  decays are averaged over longitudinal position, and the vertical line at $300~\micm$ represents a conservative estimate of the separation required to create isolated clusters in a silicon strip detector. 
}
\label{fig:tracking}
\end{figure}

Although the dark photon search does not rely on reconstructing the decaying particle's mass, measurement of charged track momenta is a second important goal.  Using ATLAS SCT modules as a point of reference (see \secref{tracker}), the hit occupancy in FASER is nearly zero and the tracks are nearly straight.  If the track separation is sufficient to produce distinct strip clusters, track finding and fitting will present no technical challenges, but the combination of modest magnetic field and high momentum limits the achievable resolution. Karimaki~\cite{Karimaki:1997ff} has calculated the  expected performance of a magnetic spectrometer under extremely general assumptions. Momentum resolution scales linearly with coordinate resolution and field strength, but quadratically with the length of the detector. The dependence on the number of measurements and where they are made is more complicated, but also calculable. 

\Figref{pResolution} shows FASER's idealized momentum resolution for different energies and numbers of sensor planes, assuming perfect alignment. The momentum resolution naturally degrades as the curvature of the track ($\rho$) decreases. The track stiffness ($K \equiv 1/\rho$) can be measured with roughly Gaussian errors.  As $K$ becomes consistent (within errors) with zero, the sign of the track's charge becomes indeterminate and only a lower limit on the momentum is possible.  For FASER, this estimated resolution limit is $1.5~\tev$ at $5\sigma$ and $2.5~\tev$ at $3\sigma$.  This implies that when a track is too straight to measure its curvature accurately, we will be able to set a fairly high bound on its minimum momentum.  At low energies, FASER should have excellent (few percent) momentum resolution to reject tracks below the $100~\gev$ analysis threshold.

\begin{figure}[tbp]
\centering
\includegraphics[width=0.38\textwidth]{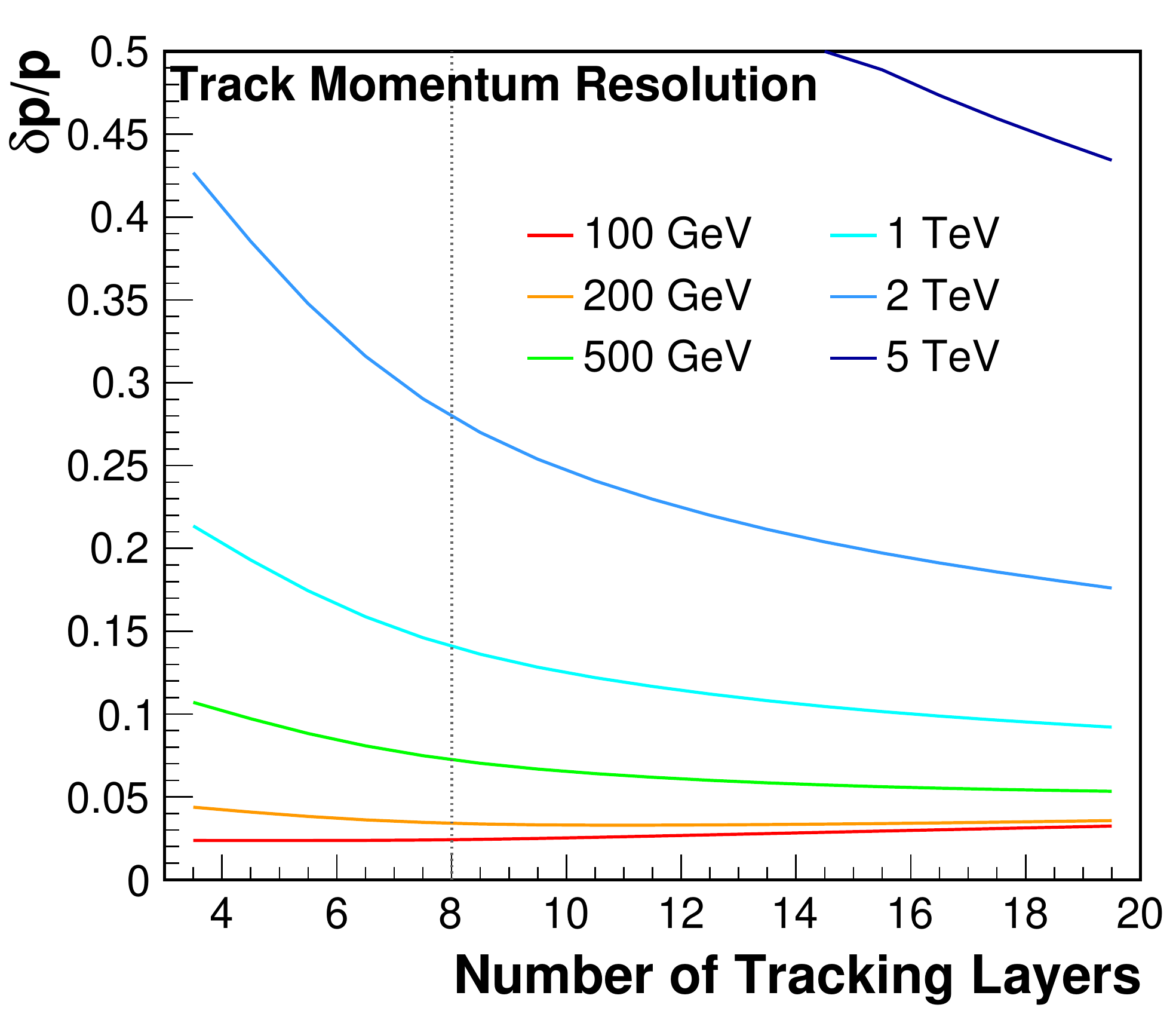} 
\caption{Estimated fractional momentum resolution ($\sigma_p/p$) as a function of the number of tracker planes for six different momenta, assuming the nominal coordinate resolution, tracker length, and magnetic field, and the theoretically optimal arrangement. FASER's choice of eight tracker planes represents a reasonable trade-off between performance, cost, and complexity.}
\label{fig:pResolution}
\end{figure}

\subsection{Calorimeter}
\label{sec:perfCalorimeter}

FASER's dark photon decay signal consists of extremely high-energy electrons, while the dominant event rate is extremely high-energy (entering) muons. To demonstrate compelling evidence of new physics, it will be important to independently establish the presence of hundreds of GeV of electromagnetic energy. Extremely precise energy resolution is not essential, but containment of TeV-energy showers requires a depth of at least 25 radiation lengths. The LHCb ECAL modules (see \secref{calorimeter}) are a robust and economical design that would be well-suited to most of our needs.  Increased leakage at FASER energies would likely degrade the nominal energy resolution ($\sigma_E/E \approx \frac{10\%}{\sqrt{E}} \oplus 1\%$) only slightly. 

For ALPs predominantly coupled to two photons, the signature is two, highly collimated photons; typical separations are shown in the right panel of \figref{tracking}.  This search would therefore benefit from a pre-shower detector of 1--2 radiation lengths depth, with the finest possible granularity, to convert and spatially resolve the photons. The possibility of adapting the silicon strip modules used in the tracker for this purpose is under study. 

\subsection{Geant4 Simulation}
\label{sec:perfSimulation}

A GEANT4-based~\cite{Agostinelli:2002hh} simulation has been developed to model the integration of detector components and their response to the signal. All sensitive elements of the detector (trigger/veto, tracker, pre-shower detector, and calorimeter) are represented. A simplified (uniform dipole) magnetic field model is used, pending a full calculation based on the magnet design.  Dark photon decays are generated with the correct kinematics, assuming uniform solid-angle and $\log p_{A'}$ distributions (which can be re-weighted in subsequent analysis). A clustering algorithm has been developed and tested to validate analytic estimates of the two-track separation efficiency. Track-finding and reconstruction algorithms using the ATLAS-derived ACTS framework~\cite{Gumpert:2017wrm} are under development.

\subsection{Signal Efficiency}
\label{sec:perfEfficiency}

The expected tracker resolution and magnetic field will make it possible to realize FASER's excellent discovery potential for new physics after applying experimental selections. In the following, we illustrate this by an analytical estimate of the signal efficiency to detect a dark photon decaying to two charged particles within FASER's decay volume. With the proposed detector we would not be able to measure the mass or the decay position of the LLP, due to the very large energies and small opening angle of the decay particles. However, to select the signal, we require that the reconstructed particles be consistent with originating from a common decay point in the decay volume. This requirement should be 100\% efficient for signal events. 

In addition, we require the dark photon decay products to (1) be completely enclosed in the tracker within a radius $R=10~\cm$, and (2) be separated by more than $\delta=0.3~\mm$ in the bending plane at the tracking stations. We consider two possible selection criteria: a \textit{loose} requirement, in which we require the tracks to be separated sufficiently in the second and third stations only, and a \textit{tight} requirement, in which we require the tracks to be separated in all three tracking stations. The signal efficiency as a function of dark photon energy and vertex position is shown in the left panel of \figref{Efficiency}. We can see that high-energy events decaying at the far end of the decay volume have a reduced signal efficiency, which is further reduced for the tight selection. There is also a slight loss of efficiency for low-energy ($< 500~\gev$) events, where the decay products can be swept outside the detector by the magnetic field. Despite these effects, the signal efficiency has only a limited impact on the sensitivity reach, which is shown in the central panel of \figref{Efficiency}.

\begin{figure}[tbp]
\centering
\includegraphics[width=0.335\textwidth]{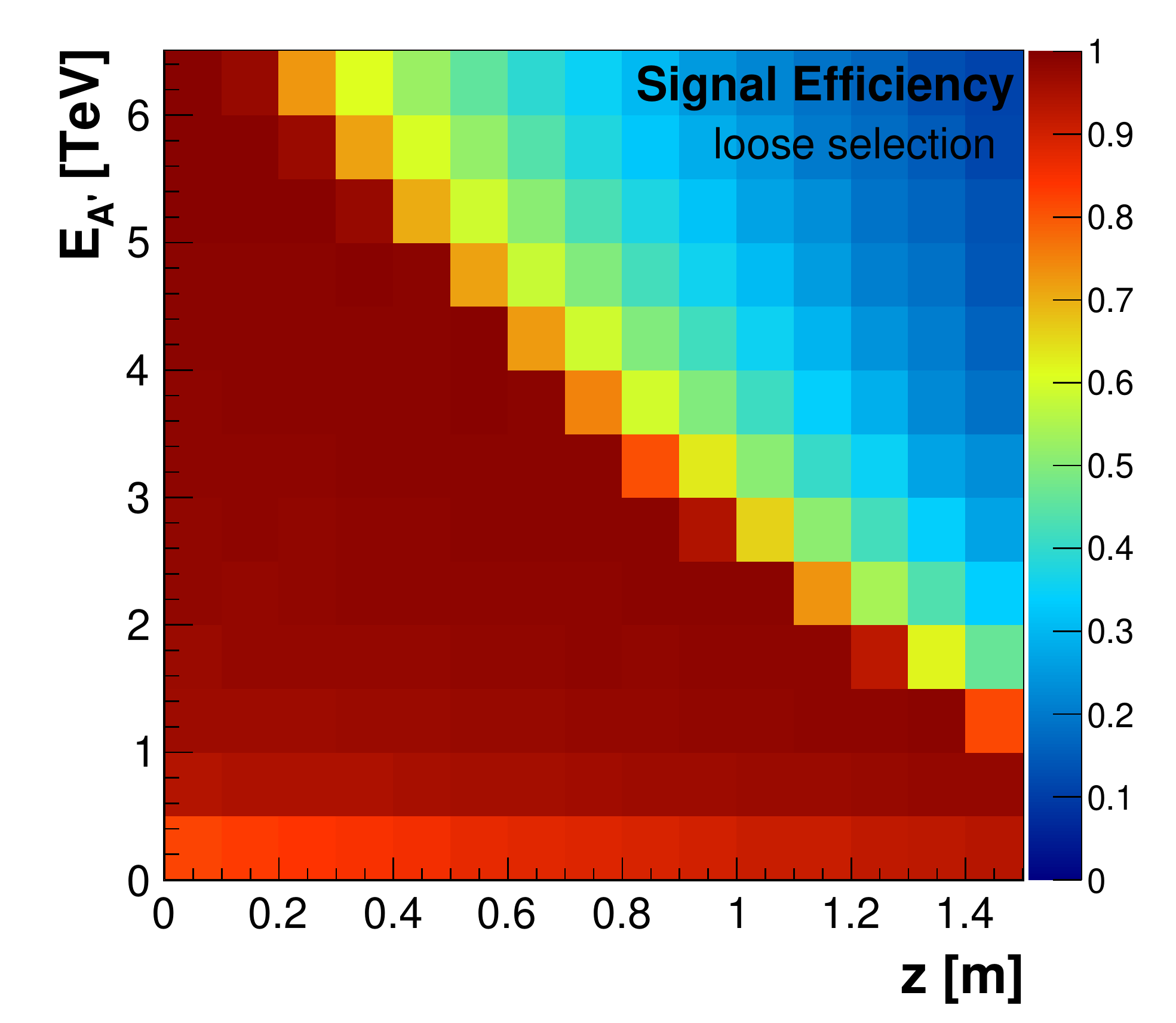} 
\includegraphics[width=0.30\textwidth]{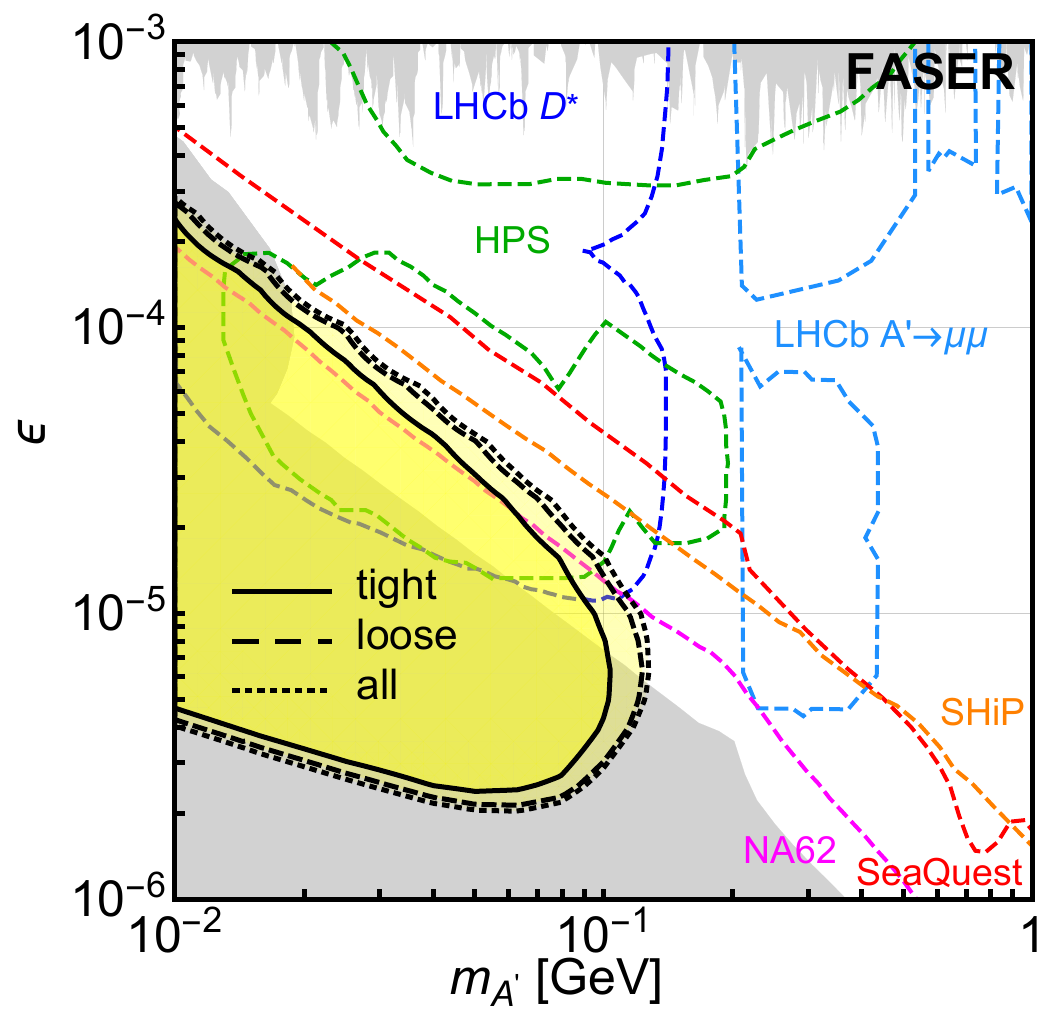} 
\includegraphics[width=0.335\textwidth]{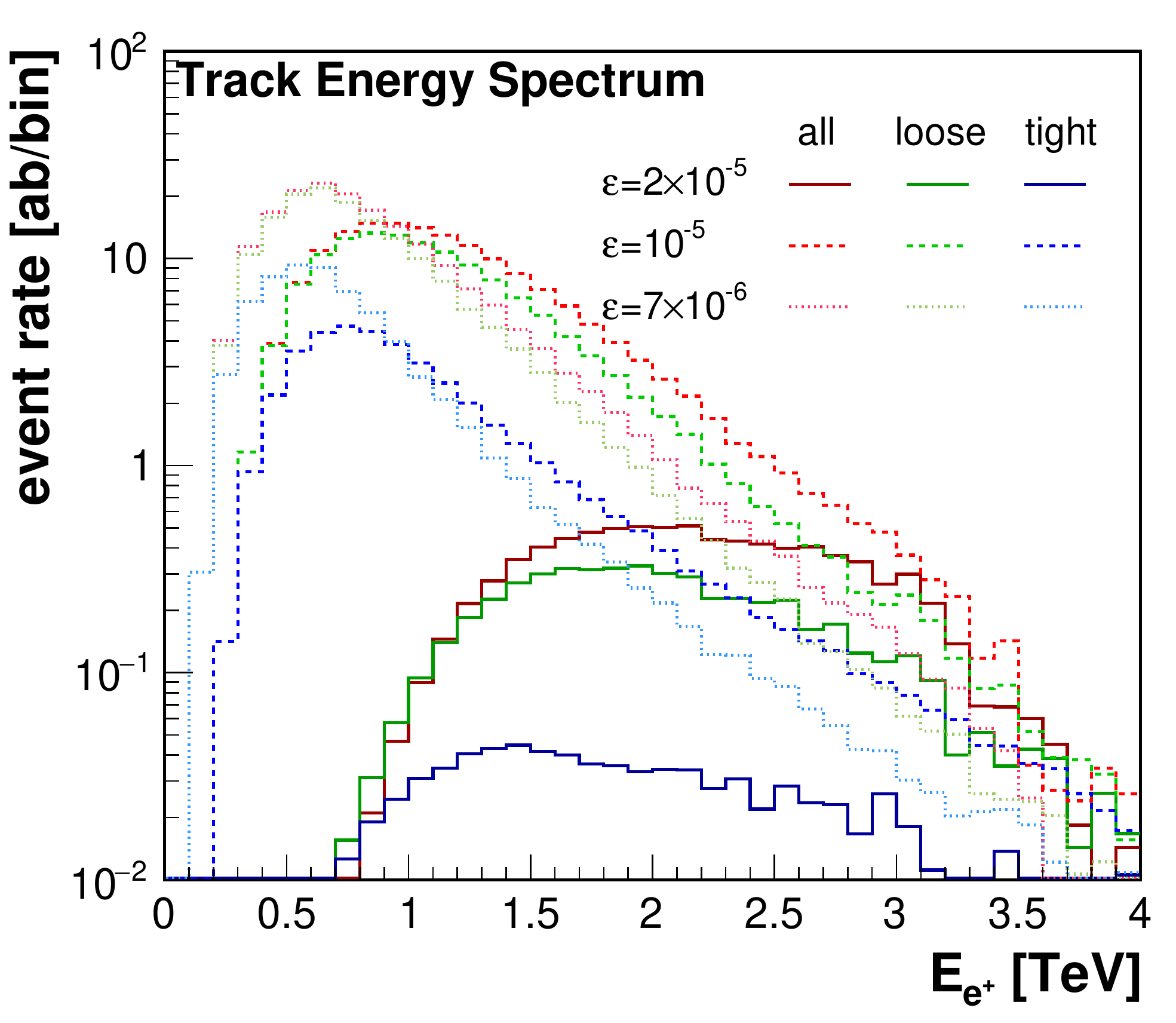} 
\caption{Left: Signal efficiency for the loose selection criterion as a function of dark photon energy and the decay's longitudinal position, averaged over the transverse position, for the dark photon benchmark point $m_{A'}=100~\mev$ and $\epsilon=10^{-5}$. Center: FASER dark photon reach without signal efficiencies (dotted), with loose selection cuts (dashed), and tight selection cuts (solid).  The ``all'' and ``loose'' curves are almost indistinguishable.  Right: Energy spectrum of dark photon decay products in FASER for $m_{A'}=100~\mev$ and $\epsilon=2\times 10^{-5}$ (solid), $\epsilon=10^{-5}$ (dashed) and $\epsilon=0.7\times 10^{-5}$ (dotted). We show the spectrum for all dark photons decaying in FASER (red), and those passing the loose (green) and tight (blue) selection cuts. 
}
\label{fig:Efficiency}
\end{figure}

In the right panel of \figref{Efficiency} we show the energy spectrum of dark photon decay products in FASER for $m_{A'}=100~\mev$ and three values of the kinetic mixing parameter $\epsilon=2\times 10^{-5}$, $1\times 10^{-5}$, $0.7\times 10^{-5}$. As can be seen, a softer spectrum is obtained for decreasing values of $\epsilon$. This is due to an increasing $A'$ lifetime, which results in a smaller boost factor for the $A'$s that can reach the detector before decaying. The colored lines show the expected spectra for all dark photons decaying in FASER (red), and those passing loose (green) and tight (blue) selection cuts. In particular the tight selection reduces the event rate at high energies, as the most energetic decay products cannot be separated enough before reaching the first tracking station. We can see that the tracks produced by a dark photon decay in FASER typically have energies of $\sim 1~\tev$ and above. 

A similar energy spectrum is also expected for photons produced by the decay of an ALP in FASER. A search for ALP decays into two photons can make use of the full volume of FASER in front of the calorimeter and requires a very good efficiency for separating two close-by showers. Importantly, as discussed in \secref{bkgRadiative}, the expected background of high-energy photons in FASER is very low and such events will typically be associated with a collinear charged particle(s) that will be detected in the front veto and the tracker. As a result, even two-photon events from ALP decays that will be mis-reconstructed as a single shower in the calorimeter could already be indicative of new physics.

\section{Detector Components}
\label{sec:detector}

\subsection{Magnets}
\label{sec:magnet}

\begin{figure}[tbp]
\centering
\includegraphics[width=0.57\textwidth]{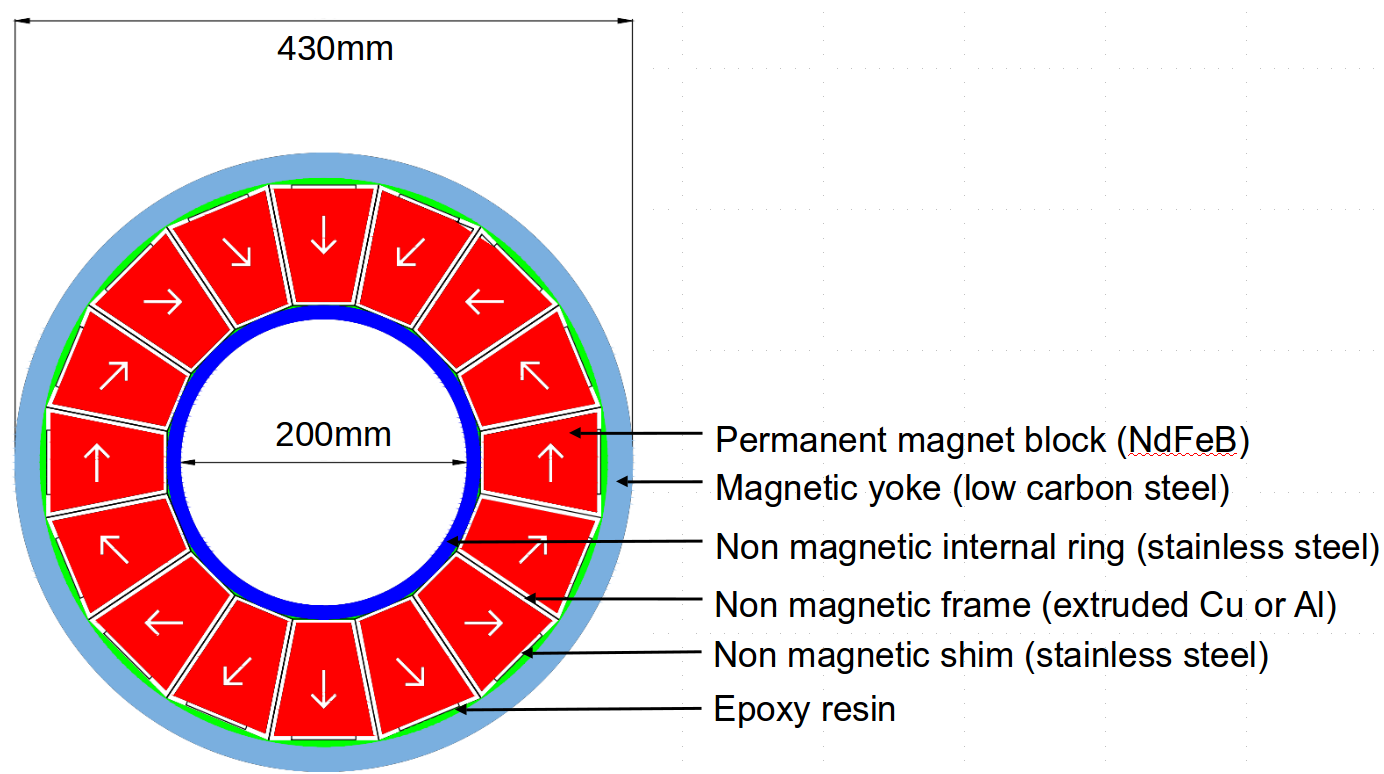} \qquad
\includegraphics[width=0.30\textwidth]{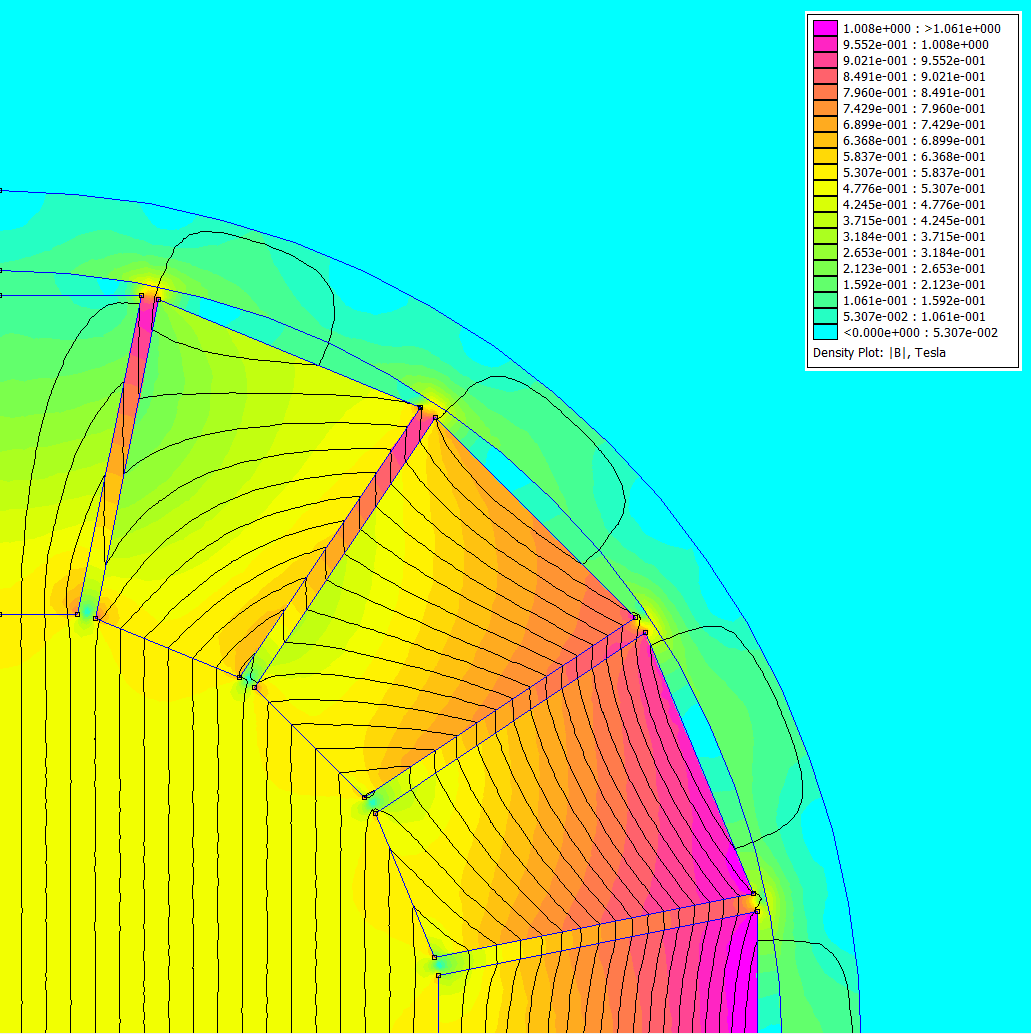} 
\caption{
Left: Halbach dipole magnet diagram (left) and field distribution (right). 
}
\label{fig:Magnet}
\end{figure}

Three dipole magnets are needed to separate energetic pairs of charged particles and to perform momentum measurements. An electromagnet would provide the strongest field, but, for the large aperture needed, is rather bulky compared to the available space and requires a significant amount of supporting infrastructure. Therefore a solution with permanent dipole magnets is preferred. The field strength of such magnets will have some dependence on the temperature, but given the high momentum of the signal particles, this is not expected to contribute significantly to experimental uncertainties.

An attractive solution is to use magnets based on a Halbach array~\cite{Halbach:1979mv} constructed from permanent magnet blocks with different magnetization directions as illustrated in \figref{Magnet}. Such a magnet design with very similar requirements was prepared by the CERN magnet group for the N-Tof experimental area, and it is expected this could be adapted for all three magnets. Using SmCo for the magnet material, a dipole field of 0.52 T can be reached.  With NdFeB the field could be as high as 0.6 T at the cost of a small increased temperature sensitivity and reduced radiation hardness, which is not expected to be a problem in the FASER location. As can be seen in the figure, such magnets are very compact compared to more traditional magnets. A one-meter-long magnet is expected to weigh about $1000~\kg$. As shown in \figref{Magnet}, the fringe fields extend out minimally radially, but, due to the large aperture, will extend out between the magnets. A minimum distance of $200~\mm$ between magnets will therefore be needed for safety reasons. These openings will be used for detector elements.

\subsection{Scintillator Trigger and Veto Layers}
\label{sec:scintillator}

\begin{figure}[tbp]
\centering
\includegraphics[width=0.62\textwidth]{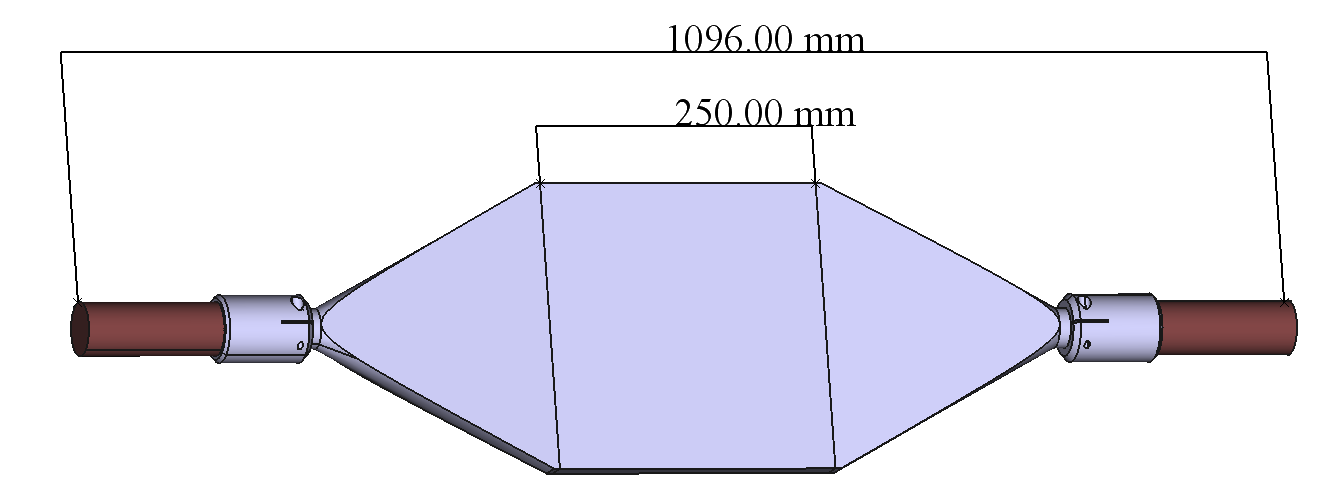} 
\caption{
Scintillator layer to be used for veto and trigger layers.
}
\label{fig:ScintillatorLayer}
\end{figure}

Four identical scintillator layers are used to provide efficient veto and trigger signals for charged particles. The basic design of each layer is shown in Fig.~\ref{fig:ScintillatorLayer} and consists of a $2~\cm$ thick, $25~\cm \times 25~\cm$ plastic scintillator connected through a light guide on each side to a PMT. The transverse size of the scintillator is larger than the magnet aperture to further ensure no charged particle can enter undetected, and the $2~\cm$ thickness is chosen to provide a high detector efficiency for these. The dual PMTs on each layer provide redundancy and ensure a very high veto efficiency for the veto layers as well as improving the timing resolution. The target efficiency for minimum ionizing particles is 99.99\% for each of the two veto stations with a timing resolution for the trigger chambers that is better than $1~\ns$. The time difference between the two trigger layers separated by $2.2~\m$ will provide rejection of signals not originating from the IP, particularly beam background coming from behind the detector. By measuring the signal amplitude, each of the trigger layers will also provide some discrimination power for the number of charged particles traversing the scintillator.

\subsection{Tracker}
\label{sec:tracker}

In the current design FASER will include 8 to 10 tracking layers:~two layers in the first tracking station, four layers in the second station, two layers in the third station, and possibly two layers inside the calorimeter as a pre-shower. Each layer consists of two single-sided silicon strip detectors with dimensions $24~\cm \times 24~\cm$, corresponding to an area of $0.06~\m^2$, which is sufficient to cover the aperture of the magnets. 

Possible candidates for the silicon strip detectors are spare modules of the SemiConductor Tracker (SCT) in the ATLAS experiment~\cite{ATLAS:1997ag, ATLAS:1997af}. The SCT has 4 cylindrical barrel layers and 18 planar endcap discs, covering $60~\m^2$. The SCT consists of 4088 independent modules which have two single-sided silicon strip detectors with a stereo angle of $40~\text{mrad}$. Each side of the modules has 768 strips with a constant pitch of $80~\text{$\mu$m}$. Figure \ref{fig:SCT} shows a barrel module with 6 on-detector ASICs per side, which are integrated into the module. These ASICs are the first stage of the detector readout, as well as setting the detector configuration. The modules for the barrel region are $6~\cm \times 12~\cm$, such that 8 modules would give 1 tracking layer in FASER. The SCT module design resolution is $17~\rm{\mu m}\times 580~\rm{\mu m}$, and the modules would be arranged in FASER such that the precision measurement is in the bending plane.
 
\begin{figure}[tbp]
\centering
\includegraphics[width=0.45\textwidth]{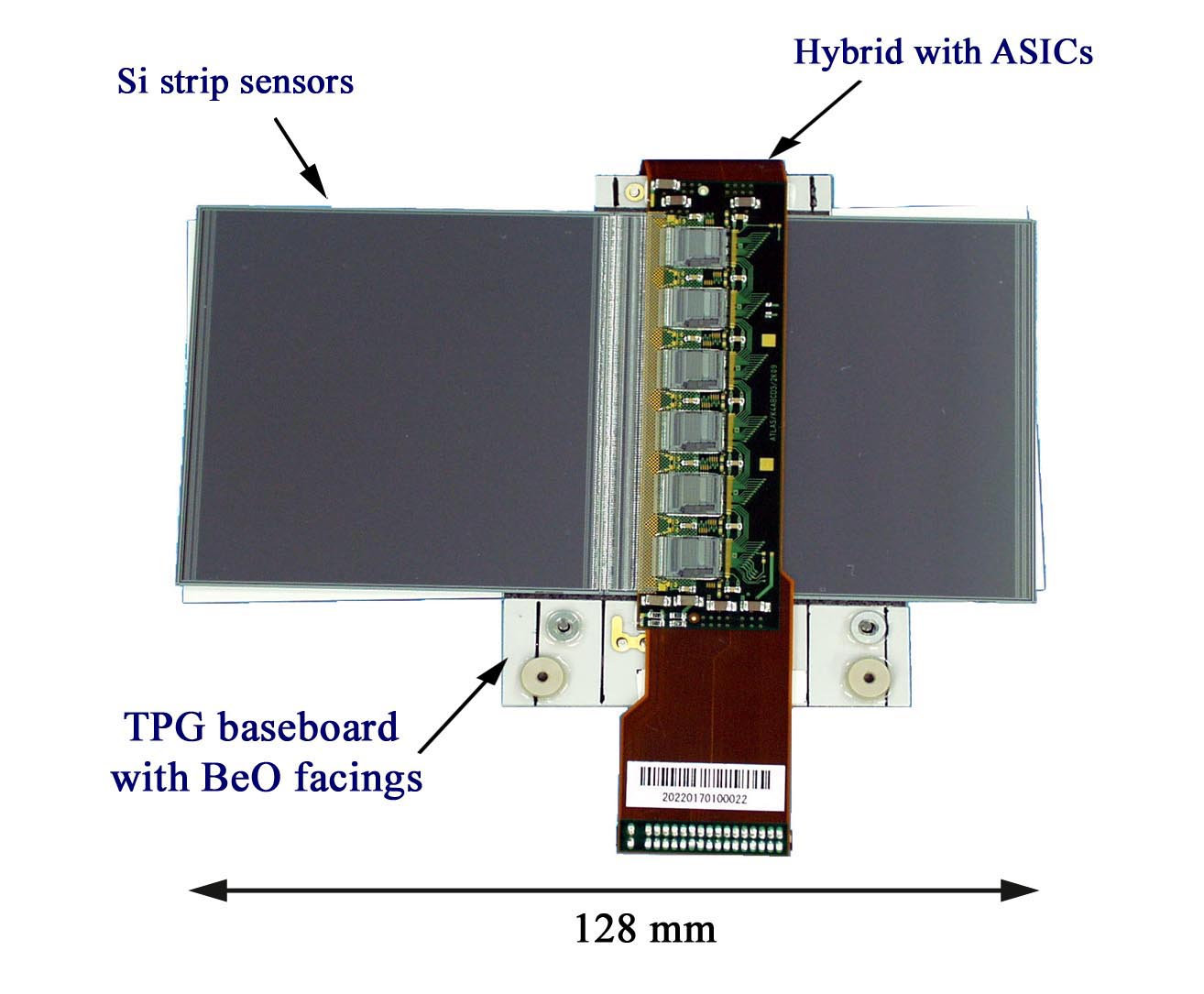}
\caption{
A picture of the barrel module of the SCT of the ATLAS experiment.
}
\label{fig:SCT}
\end{figure}

During the SCT assembly, more than 400 modules (238 modules for the barrel and 225 modules for the endcap) passing mechanical and electrical tests were kept as spare modules~\cite{Abdesselam:2006wt, Abdesselam:2007ec}. The above design could be realized using 80 of these. The ATLAS SCT power supply and interlock system could also be used in FASER. A crate for the SCT power supply houses 6 HV cards and 12 LV cards, which provide power for 48 SCT modules~\cite{Phillips:1091485}. For FASER, two SCT power supply crates would be sufficient. Given the expected low radiation levels in the TI18 tunnel, the SCT modules would be operated at room temperature using a water cooling system to cool the on-detector ASICs (5 W per module). 

An attractive solution for the tracker readout would be to use the ATLAS SCT readout system~\cite{Vickey:2006zz}. It consists of Readout Driver (ROD) VME modules, each of which is connected to one Back-of-Crate card that connects optically to up to 48 modules, therefore 2 RODs would be needed for the full FASER system. The RODs are primarily responsible for module configuration, trigger propagation, and data formatting. 

\subsection{Calorimeter}
\label{sec:calorimeter}

\begin{figure}[tbp]
\centering
\includegraphics[width=0.72\textwidth]{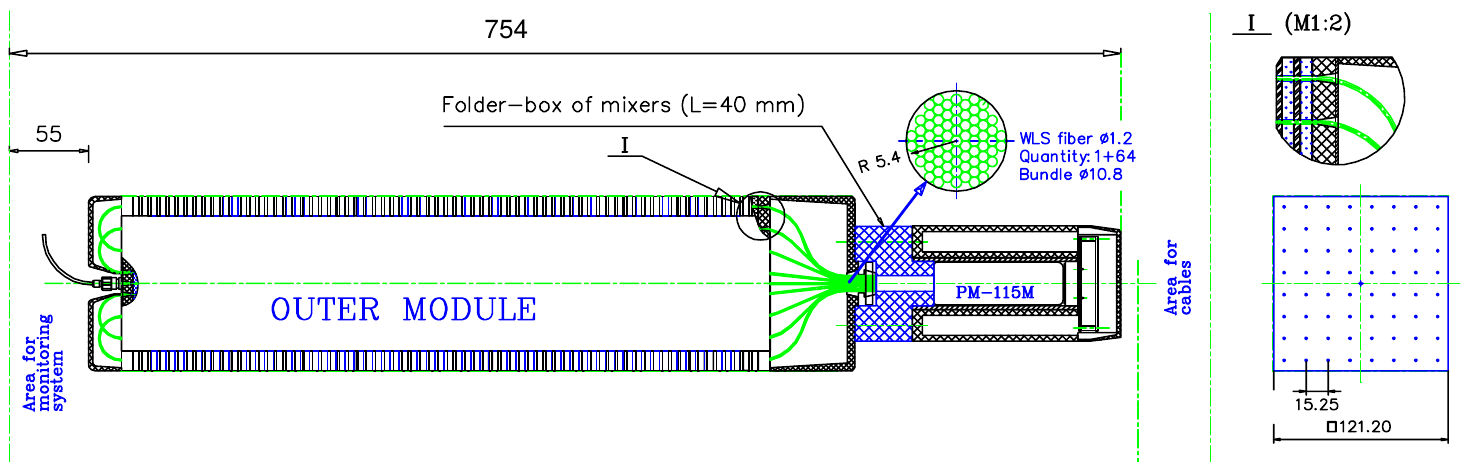} 
\caption{
Design of the LHCb outer ECAL module~\cite{LHCB:2000ab}. For FASER, the PMT might be replaced with a more compact sensor that can operate in a magnetic field.
}
\label{fig:Calorimeter}
\end{figure}

The electromagnetic calorimeter provides strong identification of high-energy electrons and photons over muons and hadrons and allows to measure their energies. Since for most signal events the $e^+e^-$ or photon pair is separated by less than a few millimeters, it is not feasible to measure the individual particle energies, and the main calorimeter requirement is therefore to measure the total electromagnetic energy with good accuracy for multi-TeV deposits in a compact detector. 

The planned calorimeter is a Shashlik-type calorimeter, as used in HERA-B and LHCb, for example, with interleaved scintillator and lead plates, and with wavelength shifting fibers penetrating the full calorimeter. The baseline is to use the same type of modules as the LHCb outer ECAL modules~\cite{LHCB:2000ab}, shown in \figref{Calorimeter}. With modules with transverse dimensions of $121.2~\mm \times 121.2~\mm$, the full FASER acceptance can be covered with just four modules. The calorimeter contains 66 layers of $2~\mm$ lead and $4~\mm$ plastic scintillator, for a total depth of 25 radiation lengths. The energy resolution for TeV deposits in such a calorimeter is expected to be around 1\%, although this will be degraded at the highest energies as 25 radiation length will not fully contain all such showers. 

To detect the presence of two electromagnetic showers separated by $300~\mu\m-2~\mm$, particularly for di-photons which are not seen in the tracker, a pre-shower detector could be placed in front of the main calorimeter. This could be constructed from a layer of tungsten or lead followed by two layers of silicon strip detectors with the strips in the two layers oriented orthogonal to each other. The thickness of the pre-shower detector is being optimized for efficiency and separation power. 

\subsection{Trigger and Readout System}
\label{sec:triggerdaq}

The detector read out will be triggered on either a coincidence in time between the two trigger scintillator layers or on a minimum amount of energy deposited in the calorimeter. The latter should provide a very efficient, but low rate selection of events with energetic electrons and photons, while the scintillator coincidence provides a trigger for signal decays to a pair of charged particles as well as a large sample of muons from the IP for alignment and calibration. 

The CAEN V1743 VME module is a candidate trigger and readout module for the scintillators and calorimeter PMTs. This module is a 16 channel, switched capacitor digitizer, which can record up to 1024 12-bit samples at 3.2 GS/s on internally generated triggers based on combinations of channels passing a discriminator threshold. Recording the full signal pulse at high precision allows for a very precise timing measurement as well as scrutinizing the details of all scintillator and calorimeter channels for non-physical anomalies in case of a possible signal. The trigger rate will be limited to less than 500 Hz to keep the dead time low. The trigger outputs will be combined in a programmable logic board, such as the CAEN V2495 module, with orbit and bunch clock signals from the LHC to align the trigger signals with IP1 collisions. The module will also generate a trigger signal for the tracker readout. No attempt will be made to combine data with the ATLAS experiment, and no signal will be exchanged between the two experiments.

The readout of the CAEN V1743 module and the tracker will be done optically to a PC located outside the LHC tunnel area. The PC will merge the two data streams and carry out additional signal processing and compression before recording to local and offline storage for data analysis. The raw output data is expected to be about 40kB/event, but compressible to less than 4kB/event, i.e. less than 2MB/s.

\subsection{Support Services}
\label{sec:support}

The best access to the TI18 tunnel is to enter the LHC at Point 1 (where the ATLAS experiment is situated) and to follow the LHC tunnel for $480~\m$.  To enter TI18, one must then cross over the LHC machine. Informal discussions with the CERN transport, civil engineering, and cryogenic teams suggest that it should be possible to transport detector components of up to about $1000~\kg$ to the location and carry them over the LHC into TI18. 

Discussions with CERN civil engineering experts suggest that excavating up to $50~\cm$ down in the tunnel floor should be possible in LS2. This is needed to have enough room for a $5~\m$ long detector to lie along the beam collision axis of the IP1 collisions.

Investigations are ongoing to find the best location to install the detector services, including a chiller for the cooling of the detector electronics, power supplies, and the readout electronics.

\section{Backgrounds}
\label{sec:bkg}

FASER's signal is multiple coincident, collimated particles of very high energy ($E \ge 100~\gev$). Muons and neutrinos are the only SM particles that can transport such energy through hundreds of meters of material between the IP and FASER. The CERN Sources, Targets, and Interactions (STI) group have computed muon fluxes at the FASER location using a FLUKA simulation. These muon fluxes, in turn, allow estimation of the rate and energy spectrum of muon-associated radiative processes near the detector. FASER-specific neutrino flux simulations will be completed in one to two months; in the interim, previous calculations of LHC neutrino fluxes are used to estimate neutrino-induced backgrounds.

To complement and validate calculated background estimates, an emulsion detector and a battery-operated radiation monitor (BatMon) began collecting data at the FASER site in June 2018.  These will provide the first {\em in situ} measurements.  

\subsection{FLUKA Simulation}
\label{sec:bkgFluka}

Recently a study from the CERN STI group~\cite{FLUKAstudy} using the FLUKA simulation program~\cite{Ferrari:2005zk,Bohlen:2014buj} was completed to assess backgrounds and the radiation level in the FASER location. The study uses a detailed geometry of the LHC and TI18 tunnels and includes the effects of the LHC infrastructure (magnetic fields, absorbers), the rock between the IP and FASER, and realistic machine optics.  Backgrounds from three sources were considered:
\begin{itemize}[leftmargin=0.16in]
\setlength{\itemsep}{-0.03in}
\item Particles produced at the IP coming directly into the FASER detector.
\item Showers initiated by protons hitting the beam pipe close to the FASER location (in the dispersion suppressor region of the LHC). These originate from off-momentum (and therefore off-orbit) protons following diffractive processes at the ATLAS IP.
\item Beam-gas interactions in beam-2 (the beam passing FASER in the direction of the ATLAS IP), which can lead to particles entering FASER without passing through any rock. 
\end{itemize}
The results show that muons are the only high-energy ($>100$~GeV) particles entering FASER from the IP, with an expected rate of 70~Hz (for the expected Run 3 conditions with a peak luminosity of $\approx 2 \times 10^{34}~\cm^{-2}~\s^{-1}$). The study shows that no high-energy particles are expected to enter FASER from proton showers in the dispersion suppressor or from beam-gas interactions.

The radiation level expected at the FASER location is very low due to the dispersion function in the LHC cell closest to FASER (cell 12).  Simulations and measurements show that the radiation level in neighboring cells (50~m upstream and downstream from FASER) are orders of magnitude larger, as can be seen in \figref{radiationCellPlot}. FLUKA predicts that the radiation level at the FASER location from proton showers in the dispersion suppressor is less than $4 \times 10^{-3}$ Gy per year or equivalently less than $4 \times 10^{7}$ 1 MeV neutron equivalent fluence per year. Beam-gas interactions are not expected to contribute due to the excellent vacuum in the LHC beam pipe. Such radiation levels are not expected to be problematic for the detector components, electronics, or services to be used in FASER. 

\begin{figure}[tbp]
\centering
\includegraphics[width=0.58\textwidth]{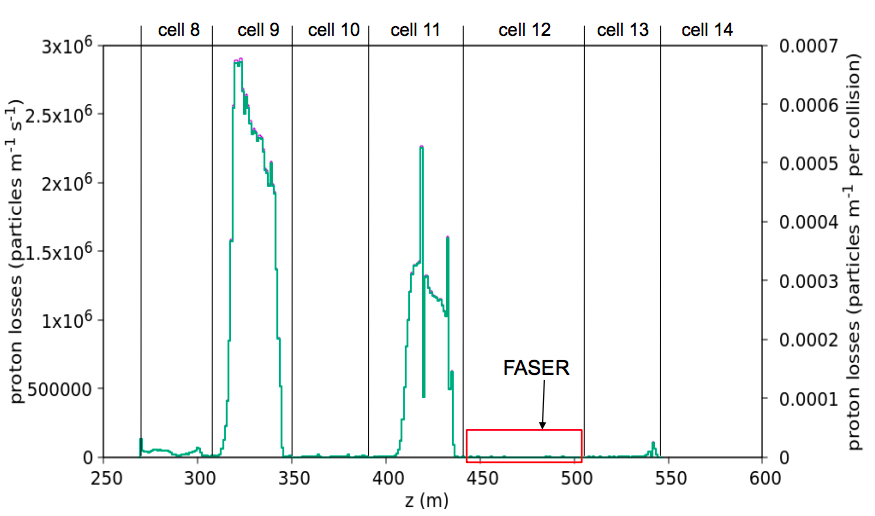} 
\caption{
Expected proton loss rate as a function of the LHC cell in the dispersion suppressor region. FASER is located in cell 12.
}
\label{fig:radiationCellPlot}
\end{figure}

\subsection{Muon-associated Radiative Processes}
\label{sec:bkgRadiative}

\tev-energy muons will produce photons and electromagnetic and hadronic showers in the rock surrounding FASER. Kinematics ensures that the scattered muon and any secondary above FASER's $100~\gev$ analysis threshold will be separated by an angle of a few mrad or less. Thus, tagging the presence of an entering muon is sufficient to differentiate these events from signal. The rates are nevertheless interesting, since they set the scale of muon rejection required and illuminate the composition of FASER's raw data.

Production of secondaries with energies above $100~\gev$ is very rare, so numerical integration is an efficient alternative to Monte Carlo. These processes are well understood, and their properties accurately parameterized~\cite{Groom:2001kq,VanGinneken:1986rf}. Rates and spectra are calculated from the FLUKA-predicted muon spectrum, assuming an exposure of $150~\fb^{-1}$, and summarized in \tableref{bkgRadiative}. Bremsstrahlung is the dominant radiative process, but most photons convert in the rock before reaching FASER. An estimated 41,000 photons with energies above $100~\gev$ will enter FASER unconverted; of these, roughly 7400 will convert in detector material before reaching the calorimeter. The muon-induced photon spectrum is sharply peaked toward lower energies (see \figref{bremPhotonSpectrum}), and all will be accompanied by the parent muon.

\begin{table}[bp]
\centering
\begin{tabular}{|l|c|}
   \hline
   \ \ {Process} & \quad {Expected Number of Events} \ \ \ \\
   \hline
   \ \  $\mu$ & 540M  \\ \hline
   \ \  $\mu + \gamma_{\text{brem}}$ & 41K \\ 
   \ \  $[\mu + (\gamma_{\text{brem}} \to e^+e^-)]$ \ \ \quad & [7.4K] \\ \hline
   \ \  $\mu + \text{EM shower}$ & 22K \\ \hline
   \ \  $\mu + \text{hadronic shower}$ & 21K \\ \hline
\end{tabular}
\caption{Expected number of events for muons and muon-induced processes that enter FASER from the direction of the IP with energy $\ge 100~\gev$ in Run 3 with integrated luminosity $150~\ifb$. One muon event occurs for every 170K bunch crossings.  The bracketed process is the subset of all $\mu + \gamma_{\text{brem}}$ events in which the photon pair converts in FASER before reaching the calorimeter.
}
\label{table:bkgRadiative}
\end{table}

\begin{figure}[tbp]
\centering
\includegraphics[width=0.58\textwidth]{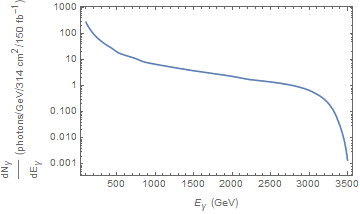}
\caption{
The calculated flux spectrum of muon-induced photons from bremsstrahlung entering FASER unconverted in an exposure of $150~\fb^{-1}$. All will be accompanied by the parent muon. 
}
\label{fig:bremPhotonSpectrum}
\end{figure}

Bremsstrahlung conversions and direct $e^+e^-$ pair production will lead to muons accompanied by electromagnetic showers, and photo-nuclear interactions will produce muons accompanied by hadronic showers. In these cases, the probability for any {\em single} particle (other than the muon) to reach FASER with $100~\gev$ or more is very small; the calculated rates are instead based on the looser requirement that the {\em total} shower energy reaching FASER (attenuated by the number of radiation or hadronic interaction lengths from the interaction, as appropriate) is over $100~\gev$. The calculation predicts that roughly $8 \times 10^4$ muons entering FASER (one in 7000) will be accompanied by additional, visible, electromagnetic or hadronic energy above $100~\gev$.  We see that, with two layers of scintillator at the front of the detector, each vetoing entering charged particles with an efficiency of 99.99\%, the backgrounds in \tableref{bkgRadiative} will all be reduced to negligible levels.

A last muon-induced background is one in which a muon first radiates a high energy photon in the rock before the detector and then decays.  We expect ${\cal O}(0.01)$ events of this kind in Run 3.

\subsection{Neutrino-induced Backgrounds}
\label{sec:bkgNeutrino}

For the large pseudorapidities characteristic of FASER with a $10~\cm$ radius, the dominant source of neutrinos is in-flight $\pi^{\pm}$ decays; heavier mesons play a less important role~\cite{Park:2011gh}. A good estimate of the high-energy neutrino flux can therefore be obtained from the forward pion spectrum, which can be convoluted with the neutrino interaction cross section to estimate the number of neutrino-induced charged current (CC) events in the detector. The result is that, requiring neutrino energies above 100 GeV (1 TeV), one expects $\sim 10$ ($\sim 0.1$) CC neutrino events per kg of detector material for $150~\fb^{-1}$ integrated luminosity~\cite{Feng:2017uoz}. Considering the small mass of the first tracking station (roughly $500~\g$) and the air in the decay volume ($60~\g$), we therefore expect at most a few $\sim 100~\gev$ CC events, and far fewer with TeV energies, where most of the signal is.  In addition, these neutrino events typically produce only one high-energy charged track, since the momentum transfer to the nucleus is form-factor suppressed, resulting in the other scattering products typically having much lower energy. For the same reason, neutral current (NC) interactions will typically only lead to  low-energy events. One therefore expects neutrino-induced backgrounds to be negligible.

\subsection{{\em In situ} Measurements}
\label{sec:bkgInSitu}

An emulsion detector was prepared and installed at the FASER location on 21 June 2018, during Technical Stop 1 (TS1). The purpose is to validate the FLUKA background estimation results. Furthermore, this measurement may pave the way for using emulsion detectors for LLP searches. Emulsion detectors are made of micro-crystals with a diameter of about 200 nm, which leads to a position resolution of 50 nm and an angular resolution of 0.35 mrad with a 200 $\mu$m-thick base~\cite{Aoki:2017spj}.  The high resolution of emulsion detectors, as well as their energy-loss ($dE/dx$) measurement capability, allows them to separate $e^+e^-$ pair signals from single electron background.  In addition, low-energy components can be rejected by their multiple Coulomb scattering inside the detector materials.

The installed detector structure is shown in the left panel of \figref{emul_str_setup}. It comprises two sections. Upstream  is a tracking section made of 10 emulsion films interleaved with 10-mm-thick Styrofoam, designed to detect two almost-parallel tracks. Each emulsion film comprises two emulsion layers ($65~\mu\m$ thick) that are poured onto both sides of a $200~\mu\m$-thick plastic base. The downstream section builds a sampling calorimeter, the so-called Emulsion Cloud Chamber (ECC), which has a repeated structure of emulsion films interleaved with 1-mm-thick or 5-mm-thick lead plates for the electromagnetic shower energy measurement. The total radiation length in the ECC is 12$X_0$. The emulsion films are vacuum-packed with a light-tight bag and enclosed in an acrylic box as shown in the right panel of \figref{emul_str_setup}. A single module, consisting of 31 emulsion films, is 173 mm wide, 124 mm high, and 210 mm thick. Additionally, two removable emulsion detectors have been placed on the front and back faces of the acrylic box to provide the possibility of a prompt check of the track density shortly after TS1. 

\begin{figure}[tbp]
\centering
\includegraphics[width=0.8\textwidth]{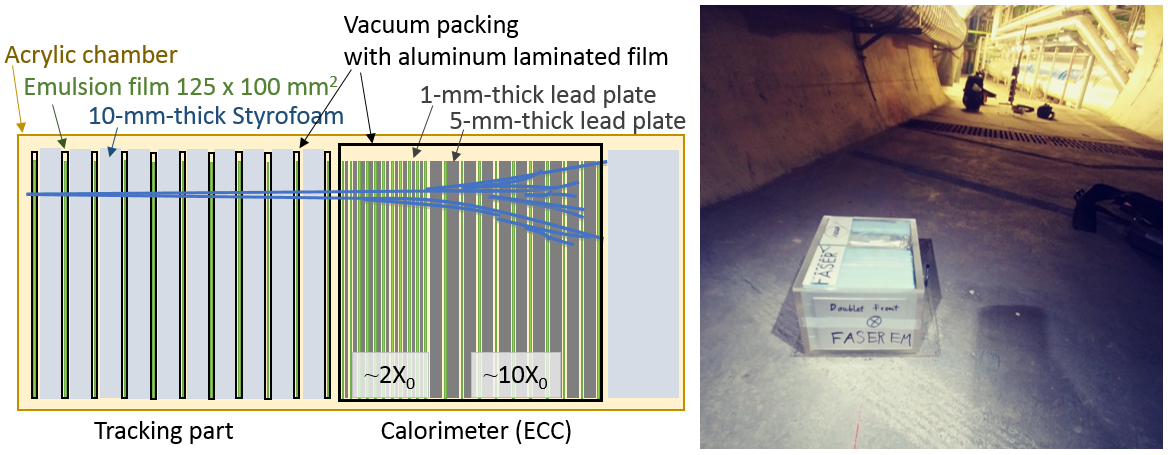}
\caption{
Left: Schematic of the emulsion detector structure. Right: The emulsion detector installed at the FASER location in TS1.
}
\label{fig:emul_str_setup}
\end{figure}

The energies of electromagnetic showers will be determined by counting the number of shower tracks in each emulsion film within a circle of radius $100~\mu\m$ centered on the shower axis~\cite{Kobayashi:2012jb}. An energy resolution of 10.6\% at 200 GeV has been found in simulations and validated by experimental data. If high-energy showers are detected in the downstream ECC section, they will be followed up to the upstream tracking section. Parallel electron tracks separated by $1~\mu\m$ can be identified and possible decay points can be estimated. 

Concerning the measurement in TS1, the additional two emulsion detectors outside the box will first be removed and analyzed before TS2. The measured track density in these layers will then inform the decision of when to remove the rest of the emulsion detector.  The data readout of the emulsion films will be performed with scanning microscopes~\cite{Ariga:2013mba,Yoshimoto:2017ufm}. The emulsion readout and reconstruction chain allows one to work with a track density of up to $10^6~\text{particles}/\cm^2$.  The track density and high-energy electromagnetic components will be measured and compared to the background prediction from the FLUKA simulations. 

To validate the FLUKA simulation results on the radiation level at the FASER location, a BatMon, commonly used in the LHC, was installed {\em in situ} in TS1. This will be read out in TS2 to measure the accumulated dose during this period.

\section{Cost, Funding, and Timeline}
\label{sec:cost}

An essential feature of FASER is its ability to do world-leading physics at a very affordable cost, thanks to the size and location of the experiment.  FASER's active volume is just $0.16~\m^3$, and the entire experiment fits in a box with dimensions $1~\m \times 1~\m \times 5~\m$.  In addition, FASER's location in TI18 is exceptionally quiet, so detector components do not need to be radiation hard, and background radiation for electronics is not a great concern.   

The most costly components of the detector are the tracker, calorimeter, and magnets. To meet the experiment's cost and schedule goals, we are actively exploring the possibility of using available spare silicon strip and calorimeter hardware from the larger LHC experiments.  As noted above, roughly 200 spare ATLAS barrel SCT modules exist; FASER would require about 80 of those.  We hope to present a formal proposal and request to the SCT Institutional Board at their September 2018 meeting. For the calorimeter, the LHCb ECAL project leader has advised us that the experiment has sufficient spares to consider a similar request from FASER. While it is important to emphasize that no commitments have been made, initial discussions are encouraging. 

If spare modules are available for the tracker and calorimeter, the largest remaining construction expense are the magnets. The CERN magnet group estimates that the magnets will cost $350~\text{kCHF}$ and require one year to construct. Combined with the smaller costs of the scintillators, PMTs, trigger/readout electronics, support services, and personnel costs such as graduate student support and collaboration travel, we estimate that FASER's total cost is 1--1.5 MCHF. A respected private research foundation has expressed interest in funding FASER at this level, and FASER will be presented as a top priority by the foundation's Program Officer for Science to its Board in late September 2018. A preliminary decision is expected then, with funding starting as early as January 2019 if CERN approves the experiment.  We also intend to seek funding from national funding agencies and other sources to support additional operations costs, as well as a possible future upgrade to FASER 2.  

The current FASER collaboration is growing and will attract even greater interest once our proposals for approval and funding are successful.  If, as hoped, we are able to use spare silicon strip and/or calorimeter hardware from existing LHC experiments, we are hopeful that interested experts from the associated institutions will choose to join us.  Also, after FASER obtains funding, we will be able to offer graduate students a unique opportunity to take part in all aspects of an LHC experiment.   

FASER will be installed in TI18 over LS2 in time to take data during Run 3. To place FASER on the beam collision axis, the floor of TI18 must be lowered by $50~\cm$; this is possible without disrupting essential services and is expected to be sufficient for any beam crossing angle planned for Run 3.  With this aggressive, but feasible, schedule, FASER will have world-leading sensitivity to a broad array of LLPs, including dark photons, ALPs, and other CP-odd scalars.

If FASER is successful, a larger version, FASER 2, with a fiducial decay volume $1~\m$ in radius and $5~\m$ in length, could be installed over LS3 and take data during the HL-LHC era. FASER 2 would require extending TI18 or widening UJ18, but would greatly extend FASER's sensitivity to more massive dark photons and probe currently uncharted territory for many other models, including dark Higgs bosons and heavy neutral leptons~\cite{Feng:2017uoz, Feng:2017vli, Batell:2017kty, Kling:2018wct, Helo:2018qej, Bauer:2018onh, Feng:2018noy,  Berlin:2018jbm}.

\section{Summary}
\label{sec:summary}

FASER will extend the LHC's physics program by searching for light, weakly coupled new particles with the potential to discover physics beyond the SM and shed light on dark matter.  If installed in LS2 and collecting data in Run 3, FASER will have unprecedented sensitivity to dark photons, other light gauge bosons, and axion-like particles with masses in the 10 MeV to GeV range.  A larger detector, FASER 2, running in the HL-LHC era, will extend this sensitivity to larger masses and will probe currently unconstrained parameter space for all renormalizable portals (dark photons, dark Higgs bosons, and heavy neutral leptons), ALPs with photon, fermion, or gluon couplings, and many other new particles. 

FASER will be placed in TI18, an existing and unused tunnel 480 m from the ATLAS IP.  To maximally intersect the beam collision axis, the floor should be lowered by 50 cm, but no other excavation is required.  FASER will run concurrently with the LHC, requiring no beam modifications and interacting with the existing experiments only in requesting luminosity information from ATLAS and bunch crossing timing information from the LHC.

At present, it appears possible that the cost of design, construction, and installation, as well as some personnel costs, for FASER will be 1--1.5 MCHF.  A private foundation has expressed interest in funding FASER at this level, with a preliminary approval decision in late September 2018 and funding beginning as early as January 2019, contingent upon CERN approval. We also intend to seek funding from national grant agencies and other sources to support additional operations costs and are actively working to increase the size of the collaboration.  

\acknowledgments

FASER gratefully acknowledges invaluable assistance from many people, including the CERN Physics Beyond Colliders study group, and in particular Mike Lamont for help in many areas; Jonathan Gall and John Osborne (civil engineering); Rhodri Jones (instrumentation); Dominique Missiaen, Tobias Dobers (survey); Caterina Bertone (transport); Francesco Cerutti, Marta Sabat\'e-Gilarte, and Andrea Tsinganis (FLUKA simulation and background characterization); Attilio Milanese and Pierre Alexandre Thonet (magnets); Christian Joram, Raphael Dumps, and Sune Jacobsen (scintillators); Dave Robinson and Steve McMahon (ATLAS SCT); Yuri Guz (LHCb calorimeters); Andreas Hoecker, Andy Lankford, Ludovico Pontecorvo, Christoph Rembser, Yoram Rozen (useful discussions).  We also thank Pierre Valentin and Salvatore Danzeca for precisely mapping out the beam collision axis in TI18 and installing the BatMon and emulsion detector in TS1. 

J.L.F. and F.K. are supported by NSF Grant No.~PHY-1620638.  J.L.F. is supported in part by Simons Investigator Award \#376204.   I.G. is supported in part by DOE grant DOE-SC0010008.  S.T. is supported in part by the Polish Ministry of Science and Higher Education under research grant 1309/MOB/IV/2015/0 and by the National Science Centre (NCN) in Poland research grant No.\ 2015-18-A-ST2-00748. 

\bibliography{FASER_LOI}

\end{document}